\pdfoutput=1
\documentclass[sigconf, nonacm]{acmart} 
\usepackage{enumitem} %
\usepackage{booktabs, siunitx}
\usepackage{balance} %
\usepackage[american]{babel} %
\usepackage{hyperref}

\usepackage{tabularx} %
\usepackage{multirow} %
\usepackage{array} %
\usepackage{subcaption}

\copyrightyear{2026}
\acmYear{2026}
\setcopyright{cc}
\setcctype{by-nc-nd}

\AtBeginDocument{%
  }

\begin{document}

\title[The Use of CT Skills, Difficulties, and Strategies of Introductory Programming Students Solving Bebras Tasks]
{The Use of Computational Thinking Skills, Difficulties, and Strategies of Introductory Programming Students Solving Bebras Tasks}

\author{Enrico Benedetti}
\affiliation{%
  \institution{Utrecht University}
  \city{Utrecht}
  \country{The Netherlands}}
\email{e.benedetti@uu.nl}
\orcid{0009-0009-2379-875X}

\author{Isaac Alpizar-Chacon}
\affiliation{%
  \institution{Utrecht University}
  \city{Utrecht}
  \country{The Netherlands}}
\email{i.alpizarchacon@uu.nl}
\orcid{0000-0002-6931-9787}

\author{Johan Jeuring}
\affiliation{%
  \institution{Utrecht University}
  \city{Utrecht}
  \country{The Netherlands}}
\email{j.t.jeuring@uu.nl}
\orcid{0000-0001-5645-7681}

\begin{CCSXML}
<ccs2012>
   <concept>
       <concept_id>10003456.10003457.10003527</concept_id>
       <concept_desc>Social and professional topics~Computing education</concept_desc>
       <concept_significance>500</concept_significance>
       </concept>
   <concept>
       <concept_id>10003456.10003457.10003527.10003528</concept_id>
       <concept_desc>Social and professional topics~Computational thinking</concept_desc>
       <concept_significance>500</concept_significance>
       </concept>
   <concept>
       <concept_id>10003456.10003457.10003527.10003540</concept_id>
       <concept_desc>Social and professional topics~Student assessment</concept_desc>
       <concept_significance>300</concept_significance>
       </concept>
 </ccs2012>
\end{CCSXML}

\ccsdesc[500]{Social and professional topics~Computing education}
\ccsdesc[500]{Social and professional topics~Computational thinking}
\ccsdesc[300]{Social and professional topics~Student assessment}

\keywords{Computational thinking, Problem solving, Bebras tasks, Introductory programming} %

\begin{abstract}

\textbf{Background and Motivation.}
Computational thinking (CT) is regarded as a fundamental skill set that everyone should learn.
Identifying when and how CT skills are used is challenging but important to inform interventions that support their development.
Previous research has examined how students, teachers, and experts apply CT skills when solving introductory computational problems.
However, the extent to which higher education students in introductory programming courses do so in depth is underexplored.

\noindent\textbf{Objectives.}
The present work addresses this gap by examining in detail how students in an introductory programming course apply CT skills when solving computational problems, the difficulties they encounter, and the strategies they employ.

\noindent\textbf{Methods.}
We collected plans and solutions to Bebras tasks---short problems introducing computer science concepts and considered effective for eliciting CT skills---in an introductory programming course for non-CS majors.
We gathered 241 submissions from 58 students across five tasks, along with post-task comments and reflections on strategies.
We analyzed the data using descriptive statistics, applied an existing coding scheme to identify CT skills, and conducted thematic analysis to identify difficulties and strategies.

\noindent\textbf{Findings.}
Student submissions varied in structure and level of detail.
The most prevalent CT skills were algorithmic thinking, abstraction, and decomposition, while evaluation and generalization appeared much less frequently.
CT skill presence was positively associated with correct answers.
Students faced challenges in four areas, including understanding the tasks and making a plan, and reported various problem-solving strategies such as reading carefully, breaking down the problem, and using pen and paper.

\noindent \textbf{Implications.}
We consolidate and extend prior research on CT skills and problem solving. 
Our findings show that university students in introductory programming apply CT skills but can struggle to solve problems systematically and explain their reasoning.
Bebras tasks create opportunities for this population to engage CT skills and could be used in future research and interventions.

\end{abstract}

\maketitle

\begingroup
\renewcommand\thefootnote{}
\footnotetext{This is the authors’ version of the work. It has been accepted for publication in the \textit{Proceedings of the ACM Conference on International Computing Education Research Vol.1, 2026}. The definitive version is published by ACM at \href{https://doi.org/10.1145/3765964.3811647}{https://doi.org/10.1145/3765964.3811647}.}
\endgroup

\section{Introduction}

Computational thinking (CT) involves solving problems by drawing on concepts and cognitive tools fundamental to computer science \cite{wingComputationalThinking2006a}.
CT builds upon research in mathematics education and problem solving \cite{wingComputationalThinkingThinking2008, disessaComputationalLiteracyBig2018, paltsModelDevelopingComputational2020a, wuIdentificationProblemSolvingTechniques2024}, notably Papert's constructionist approach to teaching with computers \cite{papertMindstormsChildrenComputers1980} and Polya's four-step problem-solving method, which consists of understanding the problem, creating a plan, carrying out the plan, and looking back on the solution \cite{polyaHowSolveIt1945a}.
CT skills include abstraction, algorithmic thinking, decomposition, evaluation, and generalization \cite{selbyRelationshipsComputationalThinking2015a, selbyComputationalThinkingDeveloping}.
Learning CT skills can help people understand and work effectively with computational approaches, which are valuable in disciplines beyond CS, such as statistics, physics, biology, and the arts~\cite{wingComputationalThinking2006a}.

Recent studies have examined how professionals and academics \cite{dejongUseComputationalThinking2024}, as well as primary school teachers \cite{baveraComputationalThinkingSkills2020}, use CT skills in problem solving.
Their findings indicate that the use of CT skills can be identified by analyzing the problem-solving process, even when problems are solved incorrectly \cite{dejongUseComputationalThinking2024}.
However, even teachers can struggle to clearly explain, step-by-step, how they solved a problem \cite{baveraComputationalThinkingSkills2020}.  
Given these results, we hypothesize that undergraduates may also face difficulties in applying CT skills and articulating their problem-solving processes.
While there is research on CT interventions in higher education \cite{hsuHowLearnHow2018a, dejongComputationalThinkingInterventions2020, lyonComputationalThinkingHigher2020}, it remains understudied how students in introductory programming courses apply CT skills, and which cognitive processes play a role.
Our objective is to understand how those students use CT skills in problem solving, what challenges they encounter, and what strategies they employ. Understanding these aspects more deeply could open new avenues for supporting the development of their CT skills.

We conducted a study in an introductory Python programming course for non-CS students, in which we collected written plans and solutions to Bebras tasks---short problems designed to promote computer science and CT \cite{dagieneItsComputationalThinking2016b, dagieneShortTasksScaffolding2022}.
Our research questions are:
\begin{enumerate}
[leftmargin=*,labelindent=\parindent,start=1,label={ \bfseries RQ\arabic*}]
\item How do higher-education students in an introductory programming course for non-CS majors apply CT skills when making plans and solving Bebras tasks?
\item What difficulties do these students experience when developing plans and solving Bebras tasks?
\item What strategies do these students use when developing plans and solving Bebras tasks?
\end{enumerate}
To answer the research questions, we identified CT skills in student submissions using a coding scheme \cite{dejongUseComputationalThinking2024} based on the conceptualization of CT by \citet{selbyComputationalThinkingDeveloping}, we performed a thematic analysis of the difficulties and strategies reported by students, and we computed descriptive statistics to examine patterns in the data.

We found that students demonstrated a range of CT skills across their submissions. Algorithmic thinking, abstraction, and decomposition were more prevalent than evaluation and generalization, though the distribution of CT skills varied by task.
Students reported various difficulties, including challenges with strategy formulation and with general problem-solving stages including task comprehension.
Some students skipped planning entirely or had no clear strategy, while others employed deliberate approaches such as reading carefully, using visual aids, or decomposing problems. %

This work contributes empirical evidence on how introductory programming students apply CT skills when solving computational tasks, extending previous research to a higher-education context.
By examining how students engage with Bebras tasks, we identify 
elements that can inform more effective interventions for CT skill development. 
Our findings suggest that Bebras tasks can be useful for developing CT skills in introductory programming~courses.

\section{Background and Related Work}
\label{sec:background}

\subsection{Computational Thinking Skills}

In 2006, Wing started promoting CT as a skill set everyone should learn and use \cite{wingComputationalThinkingThinking2008, wingComputationalThinking2006a}. Since then, researchers have discussed the importance of CT skills for students across disciplines \cite{groverComputationalThinkingK122013a}.
Consensus on a single definition of CT has not been reached \cite{shuteDemystifyingComputationalThinking2017, bonnerFormativeAssessmentComputational2021, groverComputationalThinkingK122013a}. 
However, \citet[p.~2]{ahoUbiquitySymposiumComputation2011} provides a useful general definition in line with Wing's, considering CT to be ``the thought processes involved in formulating problems so their solutions can be represented as computational steps and algorithms''.

Researchers and educators have developed multiple frameworks to describe CT and its components \cite{shuteDemystifyingComputationalThinking2017, liComputationalThinkingMore2020, groverComputationalThinkingK122013a, iste_csta_ct_2011}. 
For example, \citet{brennanNewFrameworksStudying2012} draw from classroom experiences in teaching programming through design-based learning with Scratch. 
They describe CT in terms of concepts such as loops and sequences, practices such as testing and debugging, and perspectives such as expressing one's creativity.
\citet{weintropDefiningComputationalThinking2016a} argue for a reciprocal  positive relationship between learning mathematics and science content and learning computational approaches.
Their framework was designed to support the development of lesson plans and assessments in high school STEM education.
They describe CT in terms of four practice areas: data, simulation, problem-solving, and systems thinking.
\citet{shuteDemystifyingComputationalThinking2017} review existing CT frameworks and distinguish CT from other types of thinking, such as mathematical and systems thinking. 
They propose a definition of CT which emphasizes a systematic way of reasoning about problems and includes six main facets: decomposition, abstraction, algorithms, debugging, iteration, and generalization. 
Selby and Woollard synthesize a CT framework from the most used terms in the literature;
they define CT as ``a focused approach to problem solving, incorporating thought processes that utilize abstraction, decomposition, algorithmic design, evaluation, and generalizations'' \cite[p.~5]{selbyComputationalThinkingDeveloping}. 
These five skills are briefly described in Table \ref{tab:ctskills_codebook}.
\citet{selbyRelationshipsComputationalThinking2015a} connected these skills to Bloom's taxonomy in the context of learning programming.

For our study we selected the framework described by \citet{selbyComputationalThinkingDeveloping}. 
This framework characterizes the cognitive processes and skills underlying CT. 
It does not focus on instructional practices and is not specific to programming, making it applicable to non-programming problems as well. 
Additionally, De Jong et al. \cite{dejongUseComputationalThinking2024} developed a codebook for identifying CT skills in problem-solving transcripts of Bebras tasks grounded in this framework. We use an adapted version of this codebook in our study. 
The core CT skills of the framework overlap to a large extent with skills used in related studies assessing CT skills at the K-12 \cite{atmatzidouAdvancingStudentsComputational2016, bonnerFormativeAssessmentComputational2021, luoUnderstandingStudentsComputational2020a} and undergraduate level \cite{febrianDoesEveryoneUse2018}. 
These studies draw on frameworks such as \citet{wingComputationalThinkingThinking2008} and \citet{brennanNewFrameworksStudying2012}.
Thus, the framework by \citet{selbyComputationalThinkingDeveloping} captures skills that are part of a broadly shared understanding of CT.

\subsection{Bebras Tasks and CT Skills}
\label{sec:bebras_ct}

Bebras is an initiative that aims to promote CT by presenting short problems that do not require extensive knowledge of CS or mathematics to be solved.
Bebras tasks introduce CS concepts to primary and secondary school students in an approachable way, by featuring everyday situations and animal mascots (``bebras'' means beaver in Lithuanian) \cite{dagieneShortTasksScaffolding2022}. 
The tasks use multiple-choice, open-ended, or interactive formats. 
Figure \ref{fig:t3example} presents a Bebras task used in our study.
Yearly, a community of educators develops new tasks, which are used in challenges in many countries.
During Bebras challenges, students attempt to solve as many tasks of varying difficulty as possible within a limited~time.

Bebras tasks have been used in research on CT interventions \cite{kastner-haulerLearningEnvironmentPromote2024, dagieneShortTasksScaffolding2022, zapatacaceresBebrasComputationalThinking2024} and CT assessment \cite{romangonzalezCombiningAssessmentTools2019, oliveiraQuantifyingComputationalThinking2025, lockwoodDevelopingComputationalThinking2018, corralesalvarezComputationalThinkingUniversity2025}.
The findings of \citet{lockwoodDevelopingComputationalThinking2018} suggest that Bebras tasks are generally comparable to other tests for CT in first-year undergraduate students.
\citet{dagieneDevelopingTwoDimensionalCategorization2017} have proposed a classification scheme linking CT skills to Bebras tasks.
In a mixed-methods study, \citet{izuExploringBebrasTasks2017} analyzed Bebras challenge data of more than 100,000 students across seven countries, focusing on the CT concepts targeted by the tasks and how age, gender, and estimated task difficulty influence performance. 
They also assigned CT skills \cite{barendsenConceptsK9Computer2015} to the tasks, finding that most tasks deal with algorithmic thinking or data representation. 
Araujo et al. \cite{araujoHowManyAbilities2019} used a quantitative approach to try to determine the use of five CT skills by high-school students in a dataset comprising 1,564 answers to 18 Bebras tasks, each labeled with a single CT skill. 
Factor analysis did not confirm the presence of the five skills in their dataset. This may be because each task requires multiple CT skills in combination, or because students solved the tasks differently than intended. 
Instead, two main factors, named ``evaluation ability'' %
and ``algorithmic thinking and logical reasoning,'' were identified.
Their findings suggest that fine-grained CT skills are currently difficult to measure using only the final answers to the tasks.

\subsection{The Problem-Solving Process and CT Skills}
\label{sec:ct_ps_bebras}

Problem solving is the application of intelligence to achieve a goal that is not immediately attainable \cite{polyaMathematicalDiscoveryUnderstanding1981}.
Problem-solving proficiency distinguishes experts from novices, in programming as well as in other fields \cite{robinsLearningTeachingProgramming2003}. 
Experts can use both specialized and general problem-solving strategies, while novices encounter obstacles and spend little time planning \cite{robinsLearningTeachingProgramming2003}.
Building on \citet{papertMindstormsChildrenComputers1980}, \citet{linnCognitiveConsequencesProgramming1985} has proposed that problem-solving
skills are essential when learning programming.
\citet{peaCognitiveEffectsLearning1984} have identified the ability to appropriately apply higher cognitive skills, such as planning and problem-solving heuristics, as a critical aim for education in~general. %

CT is thus relevant for novice students, as it is historically and conceptually rooted in problem solving.
Both a systematic literature review \cite{wuIdentificationProblemSolvingTechniques2024} and an empirical study \cite{wooProblemSolvedHow2022} have mapped CT skills and practices to established theories of problem solving, including Polya's heuristics, positioning CT as an evolution of problem-solving skills for the information era. 
Concretely, CT can be viewed as a form of robust problem solving built on the combination of abilities such as algorithmic thinking, abstraction, and generalization, among others \cite{iste_csta_ct_2011}. 

Given the central role of problem solving in programming and computing education, understanding how students use CT and how certain interventions promote CT requires continued research in which CT is framed as a reasoning process that manifests as observable behavior \cite{lyonComputationalThinkingHigher2020}.
Multiple literature reviews \cite{kaleliogluFrameworkComputationalThinking2016, luScopingReviewComputational2022, millsCodingComputationalThinking2024} have found that studies on CT in higher education are scarcer than in K-12. 
However, increasing research efforts in this area is important, as the CT skills of undergraduate students need to be understood and refined past the K-12 level \cite{lyonComputationalThinkingHigher2020}.

Researchers have explored how students in higher education use CT in computer science and other disciplines \cite{dejongComputationalThinkingInterventions2020, hsuHowLearnHow2018a, czerkawskiExploringIssuesComputational2015}. 
For example, \citet{millerImprovingLearningComputational2013} developed and administered exercises blending CT and creative thinking to CS1 students. The number of completed exercises was significantly associated with higher course grades and CT test scores.
\citet{berkalievInitiatingProgrammaticAssessment2014} explored whether students at various stages of an applied mathematics program are adept at using CT, with a focus on the use of computational tools, such as MATLAB.
\citet{berlandCollaborativeStrategicBoard2011} analyzed transcripts of undergraduates playing a collaborative board game and identified evidence of applications of CT when, for instance, players discussed game rules or simulated the outcomes of playing strategies. 

Most studies included in the systematic literature review by \citet{lyonComputationalThinkingHigher2020} use CT to structure courses, to develop students' CT skills, or as an assessment framework.
CT is often assessed using quantitative methods including self-reported measures, tests, and questionnaires, while qualitative methods are used less commonly.
De Jong and Jeuring \cite{dejongComputationalThinkingInterventions2020} report that only 19 out of 49 empirical papers on CT interventions in higher education actually measured CT skills. %
Overall, this body of work suggests that there is a relative lack of in-depth qualitative analyses of how undergraduates apply CT skills. 

However, several empirical studies have looked at the process of solving problems through the lens of CT skills.
\citet{lyonCharacterizingComputationalThinking2022} designed an activity in which undergraduates planned and produced models for food process engineering. 
Using thematic analysis to identify different ways of expressing CT skills and practices, they found that the activity helped students apply CT skills, with students adopting expert behaviors such as generalization.
In a study with high-school students, \citet{marwanExploringNovicesProblemSolving2024} explored problem-solving strategies in math and programming problems. They found that problem decomposition, a component of CT, was effective and highlighted the need to support the development of problem-solving skills.
Another study by \citet{maDBoxScaffoldingAlgorithmic2025} focused on scaffolding university students' problem-solving skills by targeting decomposition. They introduced DBox, a tool based on Large Language Models \cite{raihanLargeLanguageModels2025} that provides on-demand hints during the planning and implementation of a solution. 
In their experiment, participants who used DBox self-reported higher critical thinking and learning outcomes compared to the control group. %

In the context of Bebras tasks, some studies have focused beyond task design and aggregated performance measures to examine more closely how different groups of people solve problems.
De Jong et al. \cite{dejongUseComputationalThinking2024} analyzed think-aloud transcripts of computer science experts, including PhD students and university professors, solving Bebras tasks. 
They found that the CT skills assigned to tasks by the Bebras community are indeed used when reasoning about them, but using CT skills does not guarantee that the task will be answered correctly.
\citet{baveraComputationalThinkingSkills2020} collected written solutions to Bebras tasks from primary school teachers and performed content and statistical analysis.
They observed that teachers had difficulty describing and explaining the process they used to solve a task. \\ 

\noindent This work extends the current state of the art with a mixed-methods case study centered on an extensive qualitative analysis of how non-CS students in higher education apply CT skills in practice, the difficulties they encounter, and the strategies they employ.

\section{Methodology}
\label{sec:methodology}

\subsection{Participants}

\begin{table*}[!htb]
    \centering
    \caption{\boldmath Demographics of participants who completed the optional survey ($n = 30$ of $58$ total participants). Percentages are rounded to the nearest integer.}

\begin{tabular}{lrr @{\hskip 3\tabcolsep} lrr}
\toprule
Characteristic & Count & (\%) & Characteristic & Count & (\%) \\ \midrule
Main language &  &  & Academic level &  &  \\
\hspace{0.10cm}Dutch & 16 & (53) & \hspace{0.10cm}Undergraduate & 19 & (63) \\
\hspace{0.10cm}English & 7 & (23) & \hspace{0.10cm}Graduate & 11 & (37) \\
\hspace{0.10cm}Others  & 6 & (20) & Degree type &  &  \\
\hspace{0.10cm}Prefer not to say & 1 & (3) & \hspace{0.10cm}Biology, Chemistry, or Pharmacy & 6 & (20) \\
 &  &  & \hspace{0.10cm}Economics or Business Sciences & 8 & (27) \\
Gender &  &  & \hspace{0.10cm}Sustainability or Innovation Sciences & 7 & (23) \\
\hspace{0.10cm}Female & 14 & (47) & \hspace{0.10cm}Engineering & 3 & (10) \\
\hspace{0.10cm}Male & 16 & (53) & \hspace{0.10cm}History, Linguistic, or Culture & 3 & (10) \\
\hspace{0.10cm}Other or Prefer not to say & 0 & (0) & \hspace{0.10cm}Social Sciences or Psychology & 3 & (10) \\ \bottomrule
\end{tabular}

    \label{tab:demographics}
\end{table*}

\noindent The participants were students enrolled in an introductory Python programming and data science course for non-CS majors at Utrecht University, a large research university in the Netherlands.
The course had 105 students: 95 undergraduates and 10 master's students.
They were informed about the study and invited to participate through email and announcements during the course.
The study was conducted as part of a weekly course assignment. Students were required to submit at least half of these assignments throughout the course to be eligible for a resit of the final exam, if necessary.
To mitigate potential pressure arising from the second author being the course instructor, all students received a resit credit for the weekly assignment coinciding with this study, regardless of whether they submitted it.
A total of 58 students agreed to participate and submitted their plans and solutions to the tasks. 

\label{sec:demographicsurvey}

At the end of the study, we included an optional online demographic survey, designed to obtain additional information on the participants' background and previous programming experience \cite{margulieuxReviewMeasurementsUsed2019}.
Of the 58 participants, 30 completed the demographic survey; their self-reported characteristics are shown in Table \ref{tab:demographics}. 
As the course was a foundational course on programming required for various master's programs, there were approximately 5 to 10 undergraduates enrolled in a pre-master program, a transitional program for students preparing to start a master's.
Some of these students might have indicated being graduate students.
Regarding previous programming experience, for 10 respondents the course was their first exposure to Python.
Eleven respondents had little or no programming experience; 
15 had some experience with other programming languages, such as R, Stata, and Scratch, although two of these students did not feel confident in their programming abilities. 
Only four had more extensive programming experience.

\subsection{Materials}
\label{sec:tasks}

\begin{figure}[!htb]
    \centering
    \includegraphics[width=\columnwidth]{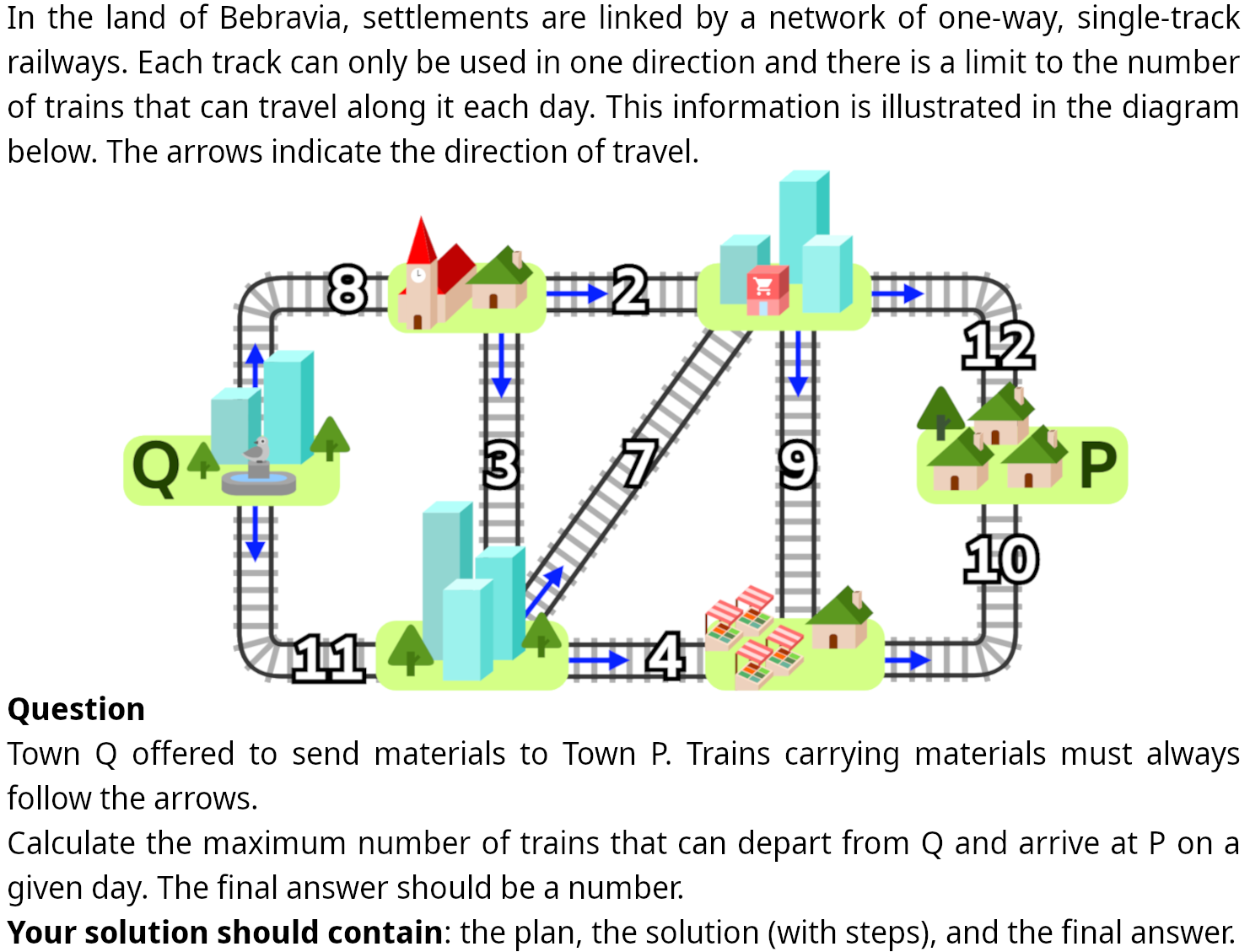}
    \caption{The third Bebras task, ``Railway Network''.}
    \label{fig:t3example}
    \Description[Task number three is shown as an example. It contains a textual description as well as an image conveying the setting of the problem.]{``Railway Network'', third Bebras task. The task contains a diagram representing connections between two cities called Q and P. They are connected by multiple one-way rails which pass through two intermediate cities or villages, where each connection has a number indicating its capacity. The problem asks to find the maximum number of trains that can go from Q to P in one day.}
\end{figure}

\noindent We used Bebras tasks because they are computational problems that can be used to examine how students apply CT skills. 
Although Bebras tasks are originally designed for students up to high-school level, they have been used to elicit problem solving and CT skills in other populations, including university students \cite{dejongUseComputationalThinking2024, baveraComputationalThinkingSkills2020, lockwoodDevelopingComputationalThinking2018}. 
The tasks are usually annotated with the major CT skills required to solve them, according to the the task authors' judgment. %

We had access to a pool of 74 English-language tasks from the 2023 and 2024 Bebras challenges in the United States and the Netherlands. 
From this pool, we selected five tasks for the assignment and two tasks for the lecture that preceded it.
Two main criteria guided our task selection: (1) appropriate challenge for undergraduate students, and (2) sufficient diversity in underlying concepts and solution approaches.

For criterion (1), we evaluated task descriptions and difficulty ratings of the 42 tasks assigned by the challenge organizers to the oldest age group, students aged 14–17. %
We initially determined that tasks classified as medium or hard (23 tasks) were appropriately challenging for undergraduates.
However, we wanted to include an easy task at the beginning to familiarize students with the requirement of explaining their problem-solving process before they moved on to more demanding problems. %
For criterion (2), we examined task descriptions and their ``explanation'' and ``background'' sections, written by task authors. %
These sections typically describe in a few paragraphs how each task can be solved, and which CS and CT concepts are involved. 
However, these sections do not always explicitly describe the required CT skills.

We first selected an easy task that had a worked-out example for a smaller-scale version of the same problem, thus providing scaffolding.
Then, we prioritized maintaining task diversity using the available information rather than relying solely on the required CT skills shown in Table \ref{tab:ctskills_per_task}.
Our aim was to expose students to a variety of topics and problems and observe how they apply CT skills. 
The tasks we used in the study covered the following concepts: sorting (T1); loops, 
variables, and randomness (T2); maximum flow (T3); sequence 
encoding and cumulative frequency (T4); and graphs and graphical 
abstractions (T5). The tasks are described below.

\begin{enumerate}
[leftmargin=*,labelindent=\parindent,start=1,label={ \bfseries T\arabic*}]
    \item ``Unload the Train'' asks to determine how many rounds of a freight train on a circular track are needed to unload boxes; a static crane must take the boxes (labeled from 1 to 10) in ascending order.
    
    \item ``Random Gift Wrap'' involves a program that prints a pattern by placing randomly-colored, -shaped, and -positioned squares and circles. Given four output patterns and the program's instructions, the task asks to  determine which pattern is impossible to obtain.
    
    \item ``Railway Network'' asks to determine the maximum number of trains that can go from one town to another in a day, following a network of one-way tracks with limited capacity. 
    Figure \ref{fig:t3example} shows the task description, as an example.
    We removed the task's candidate answers to increase its difficulty.
    This modification may shift the initial focus away from evaluation, since possible solutions are no longer given but must first be constructed, which may change the solution strategy.
    However, we believe it does not alter the overall required CT skills and it is aligned with the model solution, which focuses on computing the answer from scratch.
    
    \item ``Balls'' presents a way of encoding a sequence of blue and red balls: each ball is assigned a number based on how many blue balls follow it, plus one if it is blue itself. Even numbers in this sequence are replaced by zero, odd numbers by one. Given a sequence of zeros and ones, the task is to determine the original ball color sequence.
    
    \item ``Code Map'' asks to match the map of a fictional kingdom to its correct simplified diagram in which circles correspond to provinces and lines correspond to borders.
\end{enumerate}

Following the aforementioned criteria, we selected two additional tasks for the lecture preceding the assignment: ``Finding the Treasure'' (easy), which involved binary search in a square map, and ``Highest Sequence Score'' (hard), which involved substituting letters in a sequence to maximize the number of consecutive letters.
We presented them as worked examples to introduce students to Bebras tasks and illustrate the problem-solving process.

\subsection{Study Procedure}
\label{sec:study_procedure}

\begin{figure*}[!ht]
    \centering
    \includegraphics[width=\textwidth]{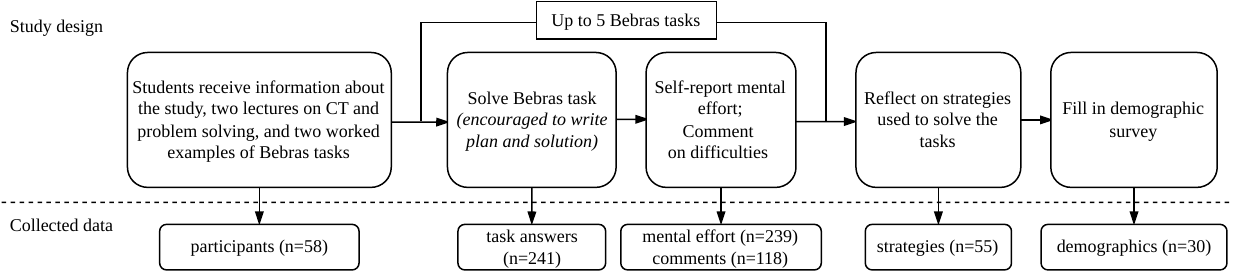}
    
    \caption{Diagram of the study design and collected data after cleaning.}
    
    \Description[Diagram of the study's procedure.]{Diagram of the study's procedure. 1. Attend the lecture on CT and planning. 2. Complete up to 5 Bebras tasks. 3. Write plan, solution, comments, and mental effort. 4. Reflect on strategies. 5. Complete demographic survey (optional).}
    \label{fig:studydesign}
\end{figure*}

\noindent Planning is an essential step in problem solving \cite{polyaHowSolveIt1945a}. 
Plans that are well-constructed help students justify and explain their reasoning and are fundamental in programming \cite{solowayLearningProgramLearning1986}. 
With these principles in mind, we asked students to solve each Bebras task in two phases: first, create a plan (analyze the problem and design an approach to solve it), 
then develop a solution (carry out the plan by performing the steps to reach an answer).

Figure \ref{fig:studydesign} summarizes what the students were expected to do in our study and shows the data we collected.
One of the authors, who was the main lecturer of the course, explained the study procedure to the students. 
Then, students received one lecture on CT and one on problem-solving strategies \cite{polyaHowSolveIt1945a}. The latter included the two Bebras worked examples. %
Following the lecture, students had one week to work on the five tasks.
The tasks were made available via a university-hosted instance of PrairieLearn, a digital learning and assessment platform \cite{prairielearn}.
Three teaching assistants checked that the assignment platform worked and that the instructions for the participants were clear.
We kept the workload consistent with other weekly assignments, so students needed to complete only the first four Bebras tasks, while the fifth was optional. 
Tasks were presented one by one, in increasing order of difficulty.
For each task, students had one text box in which to type, without restrictions on text structure.
Immediately after each task submission, students could self-report subjective cognitive load on a 7-point Likert scale ranging from very low to very high mental effort \cite{morrisonMeasuringCognitiveLoad2014}.
In another text box, they could describe any difficulties experienced while planning and solving the task.
At the end of the assignment, students were prompted to reflect on the strategies they used to solve the tasks, and were invited to complete the demographic survey.
The study was approved by the University's Ethical Review Board and conducted in February 2025. Student writings were collected on the basis of public interest for scientific research, while consent was sought for the demographic survey.

\subsection{Data Cleaning and Resulting Dataset}

We collected 247 task submissions.
We excluded three that were likely generative AI outputs and three that were empty or contained text unrelated to the tasks. 
We kept short answers to the tasks.
The final dataset consists of $241$ submissions across the five Bebras tasks, $118$ comments on difficulties, and $55$ reflections on strategies.

\subsection{Data Analysis}

\begin{table*}[!htb]
\centering
\caption{Descriptions of CT skills and subskills, slightly adapted for our study from existing literature \cite{dejongUseComputationalThinking2024, selbyComputationalThinkingDeveloping}.}
\label{tab:ctskills_codebook}
\resizebox{\textwidth}{!}{%
\begin{tabularx}{\textwidth}{ l >{\raggedright\arraybackslash}p{0.245\textwidth} >{\raggedright\arraybackslash}p{0.623\textwidth} }
\toprule
Abbr. & CT skills and subskills & Description \\ \midrule 

AB & Abstraction & The ability to decide what details of a problem are important and what details can be ignored. 
\\ 
\hspace{0.10cm} AB-REP & Choosing a representation of a system  & Choosing how to write down or draw elements of the task.   \\ 
\hspace{0.10cm} AB-RUD & Removing unnecessary details & Indicating something is not important for solving the task.  \\ 
\hspace{0.10cm} AB-SK & Spotting key elements & Indicating or mentioning an element that is important for solving the task.   \\[0.4em]

AT & Algorithmic thinking& The ability to devise explicit instructions for accomplishing tasks and creating step-by-step sets of instructions.  \\
\hspace{0.10cm}AT-CA & Creating an algorithm & Thinking about what solution algorithm one can apply, without directly applying it.   \\ 
\hspace{0.10cm}AT-EA & Executing an algorithm & Solving the task by applying the same repeated steps.  \\
\hspace{0.10cm}AT-TSR & Thinking in terms of sequences and rules  & Checking what the rules of the task are while solving the task. \\[0.4em]

DC & Decomposition & The ability to break problems down into smaller, more easily solved  parts.  \\
\hspace{0.10cm}DC-BDT & Breaking down tasks  & Stating how a problem (or parts of the problem) can be divided into actionable subtasks, or indicating what their first action to solve a task will be.  \\ 
\hspace{0.10cm}DC-INT & Dividing into subtasks with integration in mind  & Indicating how a problem can be divided into subtasks, and explicitly indicating how these can then be combined to solve the problem.  \\ 
\hspace{0.10cm}DC-CP & Thinking about problems in terms of component parts & Indicating what the different components of the problem are.\\[0.4em]

EV & Evaluation& The ability to evaluate solutions in terms of correctness, efficiency and resource utilization.\\ 
\hspace{0.10cm}EV-CCS & Checking correctness of solution  & Reflecting on the solution to the problem: is it correct?  \\ 
\hspace{0.10cm}EV-DFP & Determining fitness for purpose & Determining whether the solution fits the requirements of the task.  \\ 
\hspace{0.10cm}EV-MDR & Making decisions about good use of resources  & Reflecting on sub-solutions and the process while solving the task, are they effective or efficient enough?  \\[0.4em]

GE & Generalization & The ability to recognize elements from other knowledge contexts, to use solution components or to express them in general terms. \\
\hspace{0.10cm}GE-IPS & Identifying patterns, similarities, and connections  & Recognizing similarities to other (often CS) contexts.  \\ 
\hspace{0.10cm}GE-SNP & Solving new problems based on already-solved ones  & Applying strategies known from other (often CS) contexts.  \\ 
\hspace{0.10cm}GE-UGS & Utilizing the general solution & Mentioning or applying generalized rules such as induction or recursion to derive or justify solutions. \\ \bottomrule 
\end{tabularx}%
}
\end{table*}
To examine how students solved the Bebras tasks, we employed a mixed-methods approach combining qualitative coding and quantitative analysis.
We used several coding schemes to extract students' CT skills (RQ1), difficulties (RQ2), and problem-solving strategies (RQ3). 
We then computed descriptive statistics to characterize the data. This included correlations to explore relationships between variables of interest, such as CT skill use and task performance.

A codebook adapted from De Jong et al. \cite{dejongUseComputationalThinking2024} served as the primary guide for identifying expressions of CT skills in the Bebras task submissions, and it is shown in Table \ref{tab:ctskills_codebook}. We streamlined the subskill names and adapted descriptions for our study setting.
In addition, we created another coding scheme to record general aspects of student writings. 
Text passages were coded into: \textit{answer structure} (plan or solution), \textit{level of detail} (high-level or step-by-step), \textit{mistake}, and \textit{final answer} (correct or incorrect).
High-level passages briefly and broadly explain what students attempted, whereas step-by-step ones provide more complete and detailed descriptions.

Finally, we performed thematic analysis \cite{braunUsingThematicAnalysis2006} on the comments and reflections left by students to identify their difficulties and strategies.
We began with inductive coding, generating initial codes from the data. 
We then organized these codes into higher-level themes.
Codes related to a specific task and problem-solving phase were grouped deductively according to Polya's problem-solving~steps~\cite{polyaHowSolveIt1945a}.
For the remaining codes, we developed themes inductively.

We used NVIVO to assign and organize codes. The first and second authors performed three calibration rounds of coding CT skills. 
Each round, we sampled 15 student submissions longer than 50 characters (three per task), thus obtaining sufficient and varied content. 
We coded independently and discussed to reach agreement. 
We calculated inter-annotator agreement with Cohen’s kappa coefficient\footnote{The unit of analysis for Cohen's kappa coefficient was text characters.}
after the first calibration round, yielding 0.83 in the second round and 0.88 in the third. 
A level of agreement greater than 0.80 is considered high \cite{grahamMeasuringPromotingInterrater}. 
After the third round, the first author coded the remaining submissions and discussed them with the other authors when needed.
As for the strategies and difficulties, the first author performed the thematic analysis of post-task comments and post-assignment reflections. The resulting codes and themes were then refined through discussions with the other authors.

\section{Results}
\label{sec:results}

\noindent We first describe general information about submissions and the other data we collected, then we present the results related to the use of CT skills, difficulties, and strategies. 
Finally, we examine how students solved each of the five Bebras tasks in greater detail.
Throughout the section, we present quotes from the data to exemplify the students' use of CT skills, difficulties, and strategies. 
The quotes are left mostly unchanged, with minor edits for clarity enclosed in square brackets.
Students are assigned a numeric pseudonym to distinguish the sources of the quotes (e.g., S1).

\begin{table*}[!ht]
\centering
\caption{\boldmath General statistics of the Bebras tasks submissions: number of submissions, outcome (final answer), submission length (in words), self-reported mental effort ($1 = \textrm{very low}; 7 = \textrm{very high}$), and number of comments.}
\label{tab:generaltaskinfo}
\begin{tabular}{l r r rrrr rr rr}
\toprule
\multirow{2}{*}{Task} &
  \multirow{2}{*}{Submissions} &
  \multicolumn{4}{c}{Final answer} &
  \multicolumn{2}{c}{Words in text} &
  \multicolumn{2}{c}{Mental effort} &
  \multirow{2}{*}{Comments} \\
          &     & Correct & (\%) & Incorrect & (\%) & Mean   & SD     & Mean & SD   &     \\ \midrule
T1        & 57  & 48      & (84) & 9         & (16) & 89.47  & 49.06  & 2.64 & 1.27 & 30  \\
T2        & 56  & 16      & (29) & 40        & (71) & 80.00  & 85.58  & 3.51 & 1.28 & 32  \\
T3        & 54  & 39      & (72) & 15        & (28) & 102.63 & 80.04  & 4.04 & 1.49 & 25  \\
T4        & 55  & 40      & (73) & 15        & (27) & 121.31 & 101.54 & 4.54 & 1.49 & 24  \\
T5        & 19  & 12      & (63) & 7         & (37) & 42.32  & 49.36  & 3.78 & 1.06 & 7   \\[0.4em] 
All tasks & 241 & 155     & (64) & 86        & (36) & 93.77  & 81.28  & 3.68 & 1.52 & 118 \\ \bottomrule
\end{tabular}%

\end{table*}

\subsection{General Characteristics of Student Writings}

Table \ref{tab:generaltaskinfo} presents general statistics about the submissions ($n=241$) grouped by task.
Almost all students completed the first four tasks, while one-third completed the optional T5.
In total, 64\% of the final answers were correct. T2 was answered correctly by only 29\% of students, the lowest rate across all tasks.
Across all submissions, we identified 45 explicit mistakes (e.g., flawed reasoning).

Submission length varied greatly, ranging from texts with fewer than 10 words (28 submissions) to detailed explanations of the problem-solving process. 
On average, students wrote more for T1, T3, and T4.

\begin{table}[!htbp]
\centering
\caption{\boldmath Relative frequencies of plans and solutions in task submissions, also grouped by correctness. Percentages may not sum to $100\%$ due to rounding.}
\label{tab:plans_solution}
\begin{subtable}[b]{\columnwidth}
\centering
\caption{\boldmath All submission ($n=241$)}
\label{subtab:plan_solutions_all_grouping}
    \begin{tabular}{l *{2}{S[table-format=2.1]} S[table-format=3.1]}
        \toprule
         & {Plan} & {No plan} & {Total} \\
        \midrule
        Solution & 51.5 & 25.3 & 76.8 \\
        No solution & 7.5 & 15.8 & 23.2 \\
        Total & 58.9 & 41.1 & 100.0 \\
        \bottomrule
\end{tabular}
\end{subtable}%
\vspace{1em}

\begin{subtable}[b]{\columnwidth}
\centering
\caption{\boldmath Correct submissions ($n=155$)}
\label{subtab:plan_solutions_correct_grouping}
    \begin{tabular}{l *{2}{S[table-format=2.1]} S[table-format=3.1]}
        \toprule
         & {Plan} & {No plan} & {Total} \\
        \midrule
        Solution & 54.2 & 29.7 & 83.9 \\
        No solution & 7.7 & 8.4 & 16.1 \\
        Total & 61.9 & 38.1 & 100.0 \\
        \bottomrule
\end{tabular}
\end{subtable}%
\vspace{1em}
\begin{subtable}[b]{\columnwidth}
\centering
\caption{\boldmath Incorrect submissions ($n=86$)}
\label{subtab:plan_solutions_incorrect_grouping}
    \begin{tabular}{l *{2}{S[table-format=2.1]} S[table-format=3.1]}
        \toprule
         & {Plan} & {No plan} & {Total} \\
        \midrule
        Solution & 46.5 & 17.4 & 64.0 \\
        No solution & 7.0 & 29.1 & 36.0 \\
        Total & 53.5 & 46.5 & 100.0 \\
        \bottomrule
\end{tabular}
\end{subtable}%
      
\end{table}

Table \ref{tab:plans_solution} shows the proportion of plans and solutions in task submissions, grouped by correctness.
Correct submissions had a higher proportion of plans, solutions, as well as both combined (Tables \ref{subtab:plan_solutions_correct_grouping} and \ref{subtab:plan_solutions_incorrect_grouping}).
Generally, as shown in Table \ref{subtab:plan_solutions_all_grouping}, 84.2\% of submissions contained at least one type of answer structure (plan, solution, or both), while 15.8\% contained only a final answer with no explanatory text.
Notably, 41.1\% of submissions lacked an explicit plan. 
This includes both submissions with only the final answer (15.8\%) and those with only solution steps (25.3\%).
In two submissions, students attempted to write a plan (e.g., starting with ``plan:'') but instead described concrete steps to solve the problem, which we classified as solutions per our coding scheme.

The level of detail of text passages was further differentiated into high-level (general or vague) or step-by-step (detailed).
Plans were predominantly high-level (74\%) rather than step-by-step (26\%), whereas solutions showed the opposite pattern, with 67\% step-by-step and 33\% high-level.

We ran Pearson correlations to investigate the linear relationship between the presence of a plan or a solution and answer correctness.
Submissions containing only the final answer were significantly correlated with an incorrect answer, ${r(239)=-.27}$, ${p<.001}$.
The presence of a plan alone was not significantly correlated with a correct answer, ${r(239)=.01}$, ${p=.829}$.
Neither was the presence of a plan and a solution at the same time, ${r(239)=.07}$, ${p=.255}$ nor the presence of a plan in general, ${r(239) = .08}$, ${p=.203}$.
Solution-only submissions were weakly correlated with correctness, ${r(239)=.14, p=.036}$.
However, the correlation was stronger when accounting for all submissions with solutions in general, including those \textit{with or without} a plan, ${r(239)=.23}$, ${p<.001}$.
Writing a plan on its own may not have been sufficient, but when combined with writing the solution steps, it could have helped reach the correct~answer.

We also computed frequency tables and correlations between participant background (gender, academic level, and prior programming experience) and task outcomes (mental effort, presence of a plan, and percentage of correct answers).
No significant or strong correlations were observed.
Demographic data were available only for 30 out of 58 participants, which may have limited the test's ability to detect meaningful relationships.

\subsection{Expressions of CT Skills in Student Writings}

\begin{table*}[!htbp]
\centering
\caption{\boldmath Occurrences of CT skills expressions in the task submissions. The skill descriptions are presented in Table \ref{tab:ctskills_codebook}. The totals of the five major skills are obtained by summing the counts of the corresponding subskills. Percentages are computed over the total number of passages that were coded as CT skills ($n=366$).}
\label{tab:ctskills}
\begin{tabular}{l rrrrr @{\hskip 1.82em} rr}
\toprule
\multirow{2}{*}{CT skills and subskills} & \multicolumn{7}{c}{Frequency in task submissions} \\ 
 & T1 & T2 & T3 & T4 & T5 & All & (\%) \\ \midrule
Abstraction (AB) & \textbf{9} & \textbf{17} & \textbf{18} & \textbf{24} & \textbf{10} & \textbf{78} & \textbf{(21.3)} \\[0.15em]  %
\hspace{0.10cm}Choosing a representation of a system (AB-REP) & 1 & 0 & 2 & 0 & 3 & 6 & (1.6) \\
\hspace{0.10cm}Removing unnecessary details (AB-RUD) & 0 & 0 & 4 & 0 & 0 & 4 & (1.1) \\
\hspace{0.10cm}Spotting key elements (AB-SK) & 8 & 17 & 12 & 24 & 7 & 68 & (18.6) \\[0.4em]  %

Algorithmic thinking (AT) & \textbf{73} & \textbf{31} & \textbf{38} & \textbf{52} & \textbf{7} & \textbf{201} & \textbf{(54.9)} \\[0.15em]  %
\hspace{0.10cm}Creating an algorithm (AT-CA) & 23 & 4 & 4 & 16 & 3 & 50 & (13.7) \\
\hspace{0.10cm}Executing an algorithm (AT-EA) & 47 & 0 & 4 & 7 & 3 & 61 & (16.7) \\
\hspace{0.10cm}Thinking in terms of sequences and rules (AT-TSR) & 3 & 27 & 30 & 29 & 1 & 90 & (24.6) \\[0.4em]  %

Decomposition (DC) & \textbf{5} & \textbf{9} & \textbf{26} & \textbf{9} & \textbf{4} & \textbf{53} & \textbf{(14.5)} \\[0.15em]  %
\hspace{0.10cm}Breaking down tasks (DC-BDT) & 5 & 8 & 19 & 9 & 4 & 45 & (12.3) \\
\hspace{0.10cm}Dividing into subtasks with integration in mind (DC-INT) & 0 & 1 & 3 & 0 & 0 & 4 & (1.1) \\
\hspace{0.10cm}Thinking about problems in terms of component parts (DC-CP) & 0 & 0 & 4 & 0 & 0 & 4 & (1.1) \\[0.4em]  %

Evaluation (EV) & \textbf{0} & \textbf{5} & \textbf{5} & \textbf{14} & \textbf{1} & \textbf{25} & \textbf{(6.8)} \\[0.15em]  %
\hspace{0.10cm}Checking correctness of solution (EV-CCS) & 0 & 4 & 3 & 13 & 1 & 21 & (5.7) \\
\hspace{0.10cm}Determining fitness for purpose (EV-DFP) & 0 & 1 & 0 & 1 & 0 & 2 & (0.5) \\
\hspace{0.10cm}Making decisions about good use of resources (EV-MDR) & 0 & 0 & 2 & 0 & 0 & 2 & (0.5) \\[0.4em]  %

Generalization (GE) & \textbf{1} & \textbf{2} & \textbf{5} & \textbf{1} & \textbf{0} & \textbf{9} & \textbf{(2.5)} \\[0.15em]  %
\hspace{0.10cm}Identifying patterns, similarities, and connections (GE-IPS) & 0 & 2 & 2 & 1 & 0 & 5 & (1.4) \\
\hspace{0.10cm}Solving new problems based on already-solved ones (GE-SNP) & 1 & 0 & 2 & 0 & 0 & 3 & (0.8) \\
\hspace{0.10cm}Utilizing the general solution (GE-UGS) & 0 & 0 & 1 & 0 & 0 & 1 & (0.3) \\[0.4em]  %
All CT skills & 88 & 64 & 92 & 100 & 22 & 366 & (100.0) \\ \bottomrule
\end{tabular}%

\end{table*}

\begin{table*}[!htbp]
\sisetup{table-format=2.1}
\setlength{\tabcolsep}{7pt}
\caption{\boldmath Percentage of task submissions containing each CT skill; 
values in parentheses denote percentages of correct and incorrect submissions containing each skill, respectively.
Percentages are based on counts from Table \ref{tab:generaltaskinfo}. 
Abbreviations for the skills are introduced in Tables \ref{tab:ctskills_codebook} and \ref{tab:ctskills}.
A dagger symbol \textdagger\ indicates that the CT skill is required for solving the task; however, for T2, this information is not mentioned by the task authors.}
\label{tab:ctskills_per_task}

\begin{tabular}{
  l 
  *{5}{ r@{\,} S[table-format=2.1] @{\ (} S[table-format=2.1] @{; } S[table-format=2.1] @{)\hspace{1.2em}}}
}
\toprule
Task & \multicolumn{4}{c}{AB} & \multicolumn{4}{c}{AT} & \multicolumn{4}{c}{DC} & \multicolumn{4}{c}{EV} & \multicolumn{4}{c}{GE} \\ 
\midrule

T1 & & 14.0 & 10.4 & 33.3 & {\textdagger} & 91.2 & 93.8 & 77.8 & {\textdagger} & 8.8 & 10.4 & 0.0 & & 0.0 & 0.0 & 0.0 & & 1.8 & 2.1 & 0.0 \\

T2 & & 30.4 & 37.5 & 27.5 & & 53.6 & 56.3 & 52.5 & & 16.1 & 37.5 & 7.5 & & 10.7 & 18.8 & 7.5 & & 3.6 & 6.3 & 2.5 \\

T3 & & 27.8 & 33.3 & 13.3 & {\textdagger} & 64.8 & 76.9 & 33.3 & {\textdagger} & 42.6 & 46.2 & 33.3 & {\textdagger} & 7.4 & 10.3 & 0.0 & & 9.3 & 7.7 & 13.3 \\

T4 & {\textdagger} & 43.6 & 52.5 & 20.0 & & 67.3 & 82.5 & 26.7 & & 18.2 & 17.5 & 20.0 & & 25.5 & 30.0 & 13.3 & & 1.8 & 2.5 & 0.0 \\

T5 & {\textdagger} & 42.1 & 58.3 & 14.3 & & 26.3 & 41.7 & 0.0 & & 21.1 & 33.3 & 0.0 & & 5.3 & 8.3 & 0.0 & & 0.0 & 0.0 & 0.0 \\[0.4em]

All tasks & & 29.9 & 33.5 & 23.3 & & 66.0 & 78.7 & 43.0 & & 21.2 & 25.8 & 12.8 & & 10.4 & 12.9 & 5.8 & & 3.7 & 3.9 & 3.5 \\ 

\bottomrule
\end{tabular}
\end{table*}

\begin{figure*}[!htbp]

    \centering
    \includegraphics[width=\textwidth]{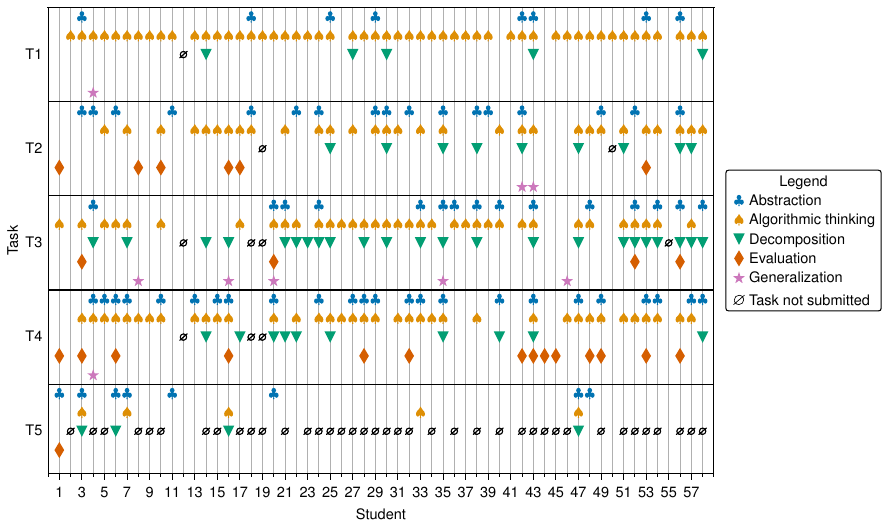}
    \caption{\boldmath Use of CT skills by each student ($n=58$) in the five tasks. Each of the five symbols denotes the presence of the corresponding CT skill; that is, at least one of its subskills was observed.}
    \label{fig:ct_skill_by_student}
     \Description[Plot of the use of CT skills by each student in the five tasks.]{For each task and each student, a symbol is present for each of the five major CT skills if the student used that skills in the task. An empty set symbol means the student did not submit any text for the task. The plot shows that the amount of CT skills used varied by student and task.}
\end{figure*}

\subsubsection{Quantitative Analysis of CT Skill Use}
Table \ref{tab:ctskills} shows the number of CT skills and subskills we identified when coding the submissions.
Algorithmic thinking was the skill coded most often (201 occurrences), while instances of generalization were the fewest (9). 
Furthermore, certain subskills were more prevalent than others within the same parent skill, such as \textit{spotting key elements} for abstraction, \textit{breaking down tasks} for decomposition, and \textit{checking the correctness of the solution} for evaluation. 

Table~\ref{tab:ctskills_per_task} shows the percentage of task submissions containing each major CT skill.
Algorithmic thinking, abstraction, and decomposition appeared more frequently, while evaluation and generalization appeared less often.
However, several tasks deviated from this general distribution.
For example, evaluation was observed more often than decomposition in T4, and abstraction appeared more than algorithmic thinking in T5.
Grouped by correctness, CT skills appeared more frequently in correct submissions than in incorrect ones, particularly the required CT skills.
The only outliers were abstraction in T1 and generalization in T3.
In some cases, the percentage difference was quite high (e.g., higher than 40\% for abstraction in T5 and algorithmic thinking in T3, T4, and T5).

Figure \ref{fig:ct_skill_by_student} presents the distribution of observed CT skills, organized by task and student. 
The vast majority of task submissions (79.7\%) contained at least one of the five skills (41.1\% had one, 27.4\% had two, 9.5\% had three, 1.7\% had four; all five did not appear together). 
In 20.3\% of submissions, we found no evidence of CT skills being used.
Submissions containing more details of the students' problem-solving processes provided more evidence of CT skills. 
In contrast, we could not identify the use of CT skills in submissions that did not provide information about plans or solution steps.
Additionally, certain tasks made it possible to observe certain CT skills more than others. 
For example, expressions of evaluation were favored in T4 compared to T3 and the other tasks.
Almost all T1 submissions demonstrated algorithmic thinking, one of the required skills.
Decomposition was required for both T1 and T3, but students used it far less frequently in T1. 

At the individual level, students applied from 0 to 12 CT skills across tasks, with a median of 6 (IQR~3‐7.75). 
In terms of variety across tasks, the number of unique CT skills used per student had a median of 3 (IQR~2-3). 
Notably, 10 students expressed exclusively a single CT skill in solving the tasks, while four students displayed the use of all CT skills and eight students showed the use of four.

Considering the relationship between the use of CT skills and answer correctness, 72.9\% of the submissions containing one or more expressions of CT skills ($n=192$) were correct. 
Conversely, only 30.6\% of the submissions without any expressions of CT skills ($n=49$) were correct.
The presence of any one of the five skills had a weak but significant correlation with correct answers, ${r(239) = .17}$, ${p=.009}$. 
Algorithmic thinking showed a moderate significant correlation with a correct answer, ${r(239)=.36}$, ${p<.001}$, and
decomposition was weakly but significantly correlated, ${r(239)=.15}$, ${p=.018}$.
Tests of the other individual skills were weakly correlated but not statistically significant.
These results suggest a positive association between the expression of CT skills and successfully solving Bebras~tasks.
\subsubsection{Qualitative Analysis of CT Skill Use}
\hfill

\noindent
\textbf{Abstraction} was observed 
across all tasks, accounting for 21.3\% of coded CT skills (Table \ref{tab:ctskills}). 
The most common subskill was \textit{spotting key elements} (ABS-SK). When using this skill, students often wrote down important details of the task description or its illustrations. For example, S22 mentioned that, in T2, ``Circles change in size but do not change colors once prepared.''
\textit{Removing unnecessary details} (AB-RUD) was observed less often, possibly because many students did not write about unimportant details of the problem. \\

\noindent \textbf{Algorithmic thinking} was the CT skill observed most often (54.9\%). This may be because all the tasks require creating or executing an algorithm or reasoning about sequences and rules. 
In some cases, students explicitly defined an algorithm, such as S40 in T3, who wrote a bulleted list at the end of their plan, ``[item 1] Identify all possible paths from Town Q to Town P by following the arrows. [item 2] Check the train limit for each track along those paths. [item 3] Find the route that allows the highest number of trains to reach Town P.'' 
In other cases, students directly wrote calculation steps or reasoned about the rules of the task, which shows the distinction between AT subskills. \\

\noindent \textbf{Decomposition} was applied by students across all five tasks (14.5\%), especially in T3.
Most types of decomposition were classified as \textit{breaking down tasks} (DC-BDT). This subskill involves deciding what actions should be taken first to solve the task. 
For example, S47's plan for T5 states, ``First draw the correct treasure map, and count the amount of borders [and] connections per node.''
\textit{Dividing into subtasks with integration in mind} (DC-INT) was identified only 4 times. 
Its description shares similarities with DC-BDT, but emphasizes reasoning about how to combine the outcomes of the subtasks. \textit{Thinking in terms of component parts} (DC-CP) was observed when students considered one area of the tracks at a time, decomposing the problem into spatial components.
For example, S57 wrote in T3 about breaking down the problem into distinct areas and combining the results, ``We need to know the transfer rate between the 4 steps [the nodes on each vertical column---see Figure \ref{fig:t3example}] and then can conclude that the minimum transfer rate between steps is the maximum throughput of the whole system.'' \\

\noindent \textbf{Evaluation} was observed in relatively few cases (6.8\%), primarily in T4.
Most students did not mention if or how they checked their solutions for correctness or efficiency. 
The most common type of evaluation occurred when students checked the correctness of their solution (EV-CCS).
We observed one instance of \textit{determining fitness for purpose} (EV-DFP) in T2, as S53 judged whether the possible solutions given by the task satisfied the requirements by process of elimination, ``as preparing a circle is a initial step which does not repeat, that means that every circle will be the same color - so answer 4 is unlikely. \ldots That leaves me with answer 1.'' 
However, their answer was incorrect, which illustrates  that despite applying appropriate CT skills and evaluation in particular, errors in solving the problem can still occur. \\

\noindent \textbf{Generalization} was rarely observed (2.5\%). This might be explained by the nature of the Bebras tasks (generalization was not strictly required) and the limited computer science background of most participants.
However, a few students did refer to computer science and math concepts. 
They mentioned ``train flows'' in T3, a reference to flow problems; functions such as ``random.choice()'' in T2; or ``bit shifts'' in T4, referring to sequences of zeros and ones.

\subsection{Difficulties Encountered by Students}
\label{sec:difficulties}

Table \ref{tab:difficulties} presents the themes and subthemes extracted from the post-task comments, in which students elaborated on what aspects of the tasks they struggled with. 
They usually mentioned one or two aspects in each comment. Of the 58 participants, 74\% (43) reported at least one difficulty.
The themes summarize the students' perceptions of the tasks, the areas in which they struggled, and the ways in which they tried to overcome difficulties.

\begin{table}[h!]
\centering
\caption{\boldmath Themes relating to difficulties of students, with counts and percentages of coded text passages ($n=125$).} %
\label{tab:difficulties}
\begin{tabular}{l r r}
\toprule
\multicolumn{1}{l}{Themes and subthemes} & \multicolumn{1}{r}{Count} & \multicolumn{1}{r}{(\%)} \\ \midrule

Perceptions of difficulty  & \textbf{44} &  \textbf{(35.2)}\\[0.15em]
 \hspace{0.10cm} Not particularly difficult & 37 & (29.6) \\
 \hspace{0.10cm} Suspicion about a task seeming too easy & 3 & (2.4) \\
 \hspace{0.10cm} Could not solve at all & 4 & (3.2) \\ [0.4em]

 Areas of difficulty  & \textbf{57} & \textbf{(45.6)} \\[0.15em]
 \hspace{0.10cm} Understanding the task & 20 & (16.0) \\ 
 \hspace{0.10cm} Devising a plan & 9 & (7.2) \\ 
 \hspace{0.10cm} Carrying out the plan & 12 & (9.6) \\
 \hspace{0.10cm} Mental load and working memory  & 16 & (12.8) \\[0.4em]

 Ways of overcoming difficulties  &  \textbf{24} & \textbf{(19.2)} \\[0.15em]
 \hspace{0.10cm} Reading carefully or looking at the images &  15 & (12.0) \\ 
 \hspace{0.10cm} Making a plan  & 1  & (0.8) \\
 \hspace{0.10cm} Using pen and paper, drawing & 8 & (6.4) \\ \bottomrule
\end{tabular}
\end{table}

Students' perceptions of task difficulty varied.
Overall, in nearly one-third of cases (29.7\%), mental effort was on the higher end of the Likert scale. %
Mental effort ratings showed an increasing trend from T1 to T4, followed by a slight drop in T5 (Table \ref{tab:generaltaskinfo}).
Students that worked on T5 experienced slightly lower mental effort on average ($3.55$) compared to students who did not submit the last task ($3.78$).
Mental effort was not found to be correlated with the presence of a plan, CT skill use, nor correctness.
As for the post-task comments, 37 of them stated that the task was not particularly difficult, with three stating that it seemed too easy. On the other hand, four comments stated not being able to solve T3 and T4 at all. 
The distribution of mental effort ratings, the insights from qualitative data, and task performance suggest that the Bebras tasks provided an appropriate level of challenge for the participants.

In the second theme, \textit{areas of difficulty}, three out of four subthemes match Polya's steps \cite{polyaHowSolveIt1945a}: understanding the problem, devising a plan, and carrying out the plan, while the last area comprises difficulties related to mental load.
The first subtheme refers to students experiencing difficulty in understanding a task, often because they perceived the description of the task to be unclear or ambiguous at first. %
Difficulties with understanding the task mostly occurred in T2 (12 out of 20), which also had the highest number of incorrect answers. %
For example, S54 mentioned being ``initially a little confused about what that printer does exactly.''
In another case, S6 was initially struggling to understand T4: ``Originally the example did not make sense. \ldots Then when I returned to the question on a different day, I understood the question differently, as intended \ldots The problem was much easier to solve, by working backwards.''

The second area of difficulty is \textit{devising a plan}. Nine students explicitly reported that planning was difficult. 
In the words of S42, ``Figuring out what the answer was wasn't hard, I put more effort into thinking about how to formulate [the] `plan' that was required.'' 

The third area, \textit{carrying out the plan}, reveals that students encountered challenges when completing the solution.
For example, S5 reflected about how the challenge of solving T3 was mitigated by being organized: ``It was tricky keeping track of all the branching possibilities, but with structured bookkeeping this wasn't too hard.''
Their statement is also related to mental load.
Sixteen students reported high mental load or overloaded working memory, for example due to trying to keep in mind many digits or manipulate sequences of symbols.

The last theme emerged from codes that expressed how students tried to overcome difficulties. 
Most comments referred to observing carefully the task description (text and images), followed by using pen and paper to create drawings and written notes.
One student observed that writing down a plan and steps helped in reducing cognitive overload, in contrast to  nine other students who found making a plan difficult.

\subsection{Strategies Used by Students}

Table \ref{tab:strategies} reports the main strategies that emerged from our thematic analysis of the students' problem-solving strategies. Table \ref{tab:strategies_small} reports how many student reflections mentioned those high-level strategies. 
The reflections were brief and focused on one or two specific aspects of the process.
The themes map onto Polya's problem-solving steps \cite{polyaHowSolveIt1945a}, similarly to how the difficulty areas did. 
The main themes are \textit{understanding the problem}, \textit{devising and carrying out the plan}, \textit{looking back and evaluating}, and, finally, \textit{no strategy}.

Students reflected on how they tried to understand the tasks.
For example, S3's strategy for understanding the problem was incremental, ``Just read carefully and try to start at the beginning and fully understand what[']s happening in every step before moving on to the next step. And eventually, understanding the smaller parts leads to understanding the whole.''

Strategies about devising a plan and carrying it out were mentioned most often. %
Overall, students mentioned  logical thinking more frequently than creative thinking. %
Other notable strategies are breaking down the problem into smaller tasks, which corresponds to decomposition in CT. This CT skill was observed often (53 occurrences). %
Making a plan was explicitly mentioned in their strategies by four students, while eight specified writing multiple or easy-to-follow steps. %
Comparing reflections with what students wrote in the task submissions shows that they applied these strategies in practice more often than they reported afterwards.

Only a few students mentioned looking back at the solution.
Those who did, retraced their solution steps to check if their answer was correct or efficient. %
Some students likely double-checked their answers and refined them. For instance, S57 revised their answers to make them clearer with the goal of ``trying to write small logical steps that are easy to follow for a future person.'' 
Another student (S51) explained writing everything down in their own words first and then shortening it.
As we did not observe many explicit instances of evaluation except in submissions for T4, this skill may have been mostly latent in the other tasks.

Finally, nine students mentioned not using any particular strategy. For example, S26 wrote, ``I didn't really have a strategy. As [At] least it didn't feel like I had one general strategy.''
S30 also reported not giving much thought to the planning phase, preferring to ``start immediately solving it, instead of really thinking of the plan in the beginning.''

\begin{table}[h!t]
\centering
\caption{\boldmath Themes relating to problem-solving strategies of students, with counts and percentages of coded text passages ($n=103$), which emerged from 55 student reflections.}
\label{tab:strategies}

\begin{tabular}{l r r }
\toprule
\multicolumn{1}{l}{Themes and subthemes} & \multicolumn{1}{r}{Count} & \multicolumn{1}{r}{(\%)} \\ \midrule
Understanding the problem  &   \textbf{26} & \textbf{(25.2)}\\[0.15em]
 \hspace{0.10cm} Paying attention to details in text or images & 6 & (5.8)\\ 
 \hspace{0.10cm} Reading carefully or multiple times  & 14 & (13.6) \\
 \hspace{0.10cm} Finding patterns or general rules in the task & 6 & (5.8)\\[0.4em]

Devising and carrying out the plan & \textbf{56} & \textbf{(54.4)}\\[0.15em]
 \hspace{0.10cm} Using logical thinking & 14 & (13.6)\\
 \hspace{0.10cm} Using creative thinking & 5 & (4.9)\\
 \hspace{0.10cm} Making a plan & 4 & (3.9)\\
\hspace{0.10cm} Writing multiple, easy-to-follow steps & 8 & (7.8) \\ 
 \hspace{0.10cm} Breaking down problem into smaller tasks & 8 & (7.8)\\
\hspace{0.10cm} Using pen and paper, drawing & 7 & (6.8)\\
\hspace{0.10cm} Working backwards & 7 & (6.8)\\
\hspace{0.10cm} Eliminating answers, process of elimination & 3 & (2.9)\\[0.4em]

Looking back and evaluating & \textbf{12} & \textbf{(11.7)} \\[0.15em]
\hspace{0.10cm} Checking efficiency, correctness & 5 & (4.9)\\
\hspace{0.10cm} Explaining, documenting for others or self	& 7 & (6.8) \\[0.4em]

No strategy &  9 & (8.7)\\

 \bottomrule
\end{tabular}
\end{table}

\begin{table}[h!t]
\centering
\caption{\boldmath Major themes relating to problem-solving strategies of students, with counts and percentages of student reflections ($n=55$). Some reflections had more than one theme.}
\label{tab:strategies_small}

\begin{tabular}{l r r }
\toprule
\multicolumn{1}{l}{Theme} & \multicolumn{1}{r}{Count} & \multicolumn{1}{r}{(\%)} \\ \midrule
Understanding the problem  &   23 & (41.8)\\
Devising and carrying out the plan & 43 & (78.2)\\
Looking back and evaluating & 11 & (20.0) \\
No strategy &  9 & (16.4)\\ \bottomrule
\end{tabular}
\end{table}

\subsection{Additional Details on CT Skills, Difficulties, and Strategies for Each of the Bebras Tasks}

\textbf{T1 ``Unload the train''} was the easiest task.
Almost all participants solved the task with a straightforward strategy: counting the loops needed to select numbers from 1 to 10 in ascending order from the given sequence. This approach was also proposed by the task authors, along with an alternative solution consisting of counting the number of inversions (pairs of elements out of their natural order) in the sequence. 
S47's plan expressed this idea in plain language: ``Count how many times the number to the right of the previous number is lower.'' 

Most students planned how to count the number of rounds, and then executed the calculations for each round.
In terms of CT skills, this approach was categorized as creating an algorithm and then executing it, making this task the one with the highest number of algorithmic thinking occurrences.
We also observed nine instances of abstraction, in which students highlighted the importance of elements of the problem, such as the sequence and the definition of ``one round''. 
In five instances, they decomposed the problem by counting the boxes that could be picked up in the first round.

Mistakes in this task included adding 1 to the count, considering the train to be traveling in the opposite direction, and picking up boxes in the wrong order.
Students reported that the task was easy in the comments, and the mean mental effort was slightly low.
Two students commented that they could easily construct an answer but that ``describing how I did it is harder'' (S23), suggesting that some students had more difficulties in writing the plan than in solving the task.
Two other students mentioned starting to think about the solution in programming language terms, ``I started thinking in a coded solution, instead of just describing what needs to be done.'' \\

\noindent \textbf{T2 ``Random Gift Wrap''} had the highest rate of incorrect answers (71\%) and mistakes (21). We hypothesize that this was because the task involves understanding pseudo-code with concepts of randomness, loops, and variable initialization, in addition to distractors. 
Students analyzed the pseudo-code of the task to derive properties of the output patterns, considered the given pattern options, and selected the pattern that could not be generated. 
The task's background information mentioned only \textit{debugging} and the concepts of loops, variables, scope, and order of execution. Debugging is sometimes considered a CT skill \cite{chendebugging2024}, but it was not included in our CT skills codebook.
This task possibly requires logic or an intuitive understanding of programming concepts.

Many students were confused by the distractors and reached the wrong conclusion.
This confusion probably prompted the high number of comments about difficulties understanding T2.
In one comment, S17 reflected: ``I was wondering if on pattern 2 circles could have been placed behind the squares. Also if the patterns should contain only 2 circles and 2 squares or they can have more - it was not very clear.''
S1 answered correctly, reasoning about the steps of the pseudo-code: ``[The answer is] 4, the last step always is positioning a square, but in [answer] 4 the squares are under the circles. Another thing is that the circles have different colors which is impossible following the steps.''
They may have read the problem more carefully or were more familiar with programming concepts.\\

\noindent Students solved \textbf{T3 ``Railway Network''} by analyzing the problem text and its diagram.
Some of them correctly pointed out, applying abstraction, that the central area is a bottleneck, as T3's model solution states. The maximum number of trains that can travel from town P to town Q in one day is equal to the sum of the capacities of the middle tracks. 
We also observed other ways to solve the task, including writing down all feasible routes and selecting the one with the maximum throughput.
In these cases, students used decomposition to break down the task by enumerating the routes and determining their capacities. They made lists, calculations, and schemas with arrows.
The task requires algorithmic thinking, decomposition, and evaluation. We observed instances of all of them, although evaluation was the most elusive ($\textrm{EV}=7.4\%$, Table \ref{tab:ctskills_per_task}). %
This was the only task in which students used the subskill \textit{thinking in terms of component parts} (DC-CP), when they divided the railway map into areas and tackled each separately.

In terms of difficulties, students commented that it was initially hard to keep track of the many possible paths and their capacity and found the task ``a bit tricky'' (S25).
S20 wrote: ``I had a lot of trouble finding a systematic approach to complete the task and write the plan. I also still don't [know] 100\% for sure that my solution is the optimal one (because I didn't do it very systematically).''
Another student found that using pen and paper ``helped immensely'' (S47).\\

\noindent In \textbf{T4 ``Balls''}, students worked through sequences of colors and binary numbers. They recognized that one way to solve it is to work backwards. 
The task authors noted that the task primarily requires abstraction. We observed this skill in 44\% of the submissions, often together with algorithmic thinking (67\%) and decomposition (18\%).
Many students documented each step at length, resulting in T4 submissions containing the most words on average.
In this task, we observed the highest number of occurrences of evaluation (14). Specifically, students checked the correctness of the solution after finding the answer by verifying that the sequence found could be translated back to the original.

Several students did not understand how the sequences were translated or found them difficult to visualize.
For example, S58 wrote, ``Yeah the fact you had to `translate' twice was a bit hard.''
Aside from rule misunderstandings, mistakes included typos or errors in otherwise correct solutions that demonstrated CT skill~use.\\

\noindent \textbf{T5 ``Code Map''}, the last, optional task, received only 19 submissions. 63\% were correct solutions. 
This task deals with the concept of graphs and requires abstraction. 
We observed this skill in 42\% of this task's submissions. Students had to spot the important elements in the map representations (nodes and edges) to find the two equivalent maps and discard the incorrect maps.
One student (S7) reported assigning letters to each province to derive and draw the map, comparing it with the given maps. 
We observed the combined use of decomposition (DC-BDT) and algorithmic thinking (AT-CA) in this submission: ``Give each province a letter to make life easier. For the province with the treasure, count how many prov[i]nces it shares a border with, eliminate the maps that don't have the treasure province bord[er]ing the many provinces.'' 
The next sentence contains an occurrence of abstraction (AB-REP and AB-SK): 
``I am labeling the provinces A-G. Going from left to right, from top to botto[m] con[si]dering the highest poin[t] of each province. 
Map 2 does not have 7 provinces, thus can be eliminated.''

This task required a relatively complex visual comparison of the maps' features. 
One student (S20) commented, ``It was slightly difficult to see if the map I drew matched the given maps, because things were moved around.'', and ``I was quickly able to think of a way of abstracting the maps and only focus on the important things, though.''

\section{Discussion}
\label{sec:discussion}

With respect to RQ1, our analysis shows that students applied CT skills when solving the tasks, with their writings containing different patterns of CT skill use.
Analyzing problem-solving traces provides information about the reasoning process and the use of CT \cite{dejongUseComputationalThinking2024}.
The skills we identified often coincided with those specified by task authors.
However, some students applied fewer, more, or different skills altogether.
Despite the removal of candidate answers from T3, the required CT skills were still applied, including evaluation, though it appeared less frequently;
students approached the problem similarly to the model solution.

Our findings align with those of similar studies on different populations \cite{baveraComputationalThinkingSkills2020, dejongUseComputationalThinking2024}. 
Consistent with \citet{baveraComputationalThinkingSkills2020}, some participants had difficulty describing how they solved a problem. 
Those participants expressed their reasoning in short sentences and general terms, or did not provide a plan or steps when solving a task, which hindered the identification of CT skills in their submissions.
De Jong et al. \cite{dejongUseComputationalThinking2024} found that experts apply CT skills when solving Bebras tasks, even when their answers are incorrect.
Likewise, our participants applied CT skills despite not reaching the correct answer, as shown in Table \ref{tab:ctskills_per_task}. 
Moreover, we found that students who wrote plans and explained their problem-solving processes expressed more CT skills and were more likely to solve the tasks correctly.
While the causal direction of these relationships cannot be established, we hypothesize that encouraging plan writing could improve performance on the tasks and the development of CT skills.
These findings and the hypothesis are consistent with the results of \citet{griggEffectsStudentStrategies2012}, which indicate that encouraging planning activities had positive effects on problem-solving for students in an introductory engineering course. 
They also highlight the benefits of understanding and restating the problem before planning.
In the context of programming, Soloway~\cite{solowayLearningProgramLearning1986} posits that learning to plan and build explanations supports learning to program, suggesting that such benefits may extend to CT skills development.

\label{sec:low_ct_skills_bebras}
Regarding specific CT skills, algorithmic thinking and decomposition were the most frequently observed, and their use was significantly correlated with answering correctly.
Algorithmic thinking and decomposition may be more readily expressed in writing for the tasks included in our study.
Conversely, evaluation and generalization were observed less often.
Combined, they appeared in fewer than 15\% of submissions. %
These two skills may have been applied tacitly, as \citet{dejongUseComputationalThinking2024} suggested. 

Another consideration is that few Bebras tasks are explicitly designed to involve evaluation and particularly generalization \cite{izuExploringBebrasTasks2017}. 
To elicit and train CT skills that are harder to observe and less prevalent in the task pool, new Bebras tasks targeting these skills should be developed, as \citet{dagieneDevelopingTwoDimensionalCategorization2017} recommended. 
Another possibility is integrating existing tasks into learning activities that emphasize underrepresented CT and problem-solving skills.
For example, students could be explicitly invited to reflect on the correctness of their solutions or to connect them with previous knowledge, by following a structured problem-solving process such as Polya's \cite{polyaHowSolveIt1945a}.
Furthermore, there are potential avenues for training CT skills through student collaboration and communication in the problem-solving process, for example by comparing steps or sharing written plans. 
In \citet{lyonCharacterizingComputationalThinking2022}, students worked in groups, and researchers observed many instances of generalization during the planning phase of their study's modeling activity.

Our results for RQ1 raise further questions about how competently undergraduates apply CT skills.
As we focused our coding efforts on identifying the occurrences of CT skills and subskills, minutely assessing the quality of CT skill application fell outside the scope of this study.
\citet{sulistiyowatiExplorationStudentsComputational2024}, for instance, used a rubric to assess how well grade 10 students applied CT skills, which points to an avenue for similar in-depth research at the undergraduate level.

RQ2 investigates the difficulties that students encountered while solving the tasks.
Self-reported measures of mental effort indicate that, overall, task difficulty increased progressively but remained moderate. 
Mental effort was not correlated with plan writing or correctness. 
Other factors and student characteristics, such as motivation and self-regulation~\cite{senRelationsPreserviceTeachers2023}, may have had a greater influence on task performance, given that the data were collected during a week-long assignment.

Although the Bebras tasks we selected were designed for high-school-level challenges, and most participants reported finding them not too difficult, many students did encounter difficulties.
Our thematic analysis of the comments students left after solving Bebras tasks contributes insights into what students struggled with.
Most commonly, students reported having difficulties with understanding tasks and their submissions show that sometimes they misunderstood task requirements.
For those students, describing the problem in their own words could be useful to improve task comprehension \cite{polyaHowSolveIt1945a, pugaleeComparisonVerbalWritten2004, zhouApplicationMetacognitivePlanning2023}.
In addition, students mentioned that it was hard to devise a plan and reason about certain tasks. 
Documenting their problem-solving process may have added to their cognitive load. 
It is possible that they were not used to combining pieces of the solution together \cite{solowayLearningProgramLearning1986}.

To address RQ3 about the strategies used by students when developing plans and solving Bebras tasks, we conducted a thematic analysis of their reflections and task-specific comments.
We linked reported strategies to established problem-solving heuristics \cite{polyaHowSolveIt1945a}, finding that most students could describe aspects of their meta-cognitive strategies for dealing with the tasks.
Their strategies for understanding the problem involve carefully reading task descriptions and studying illustrations to identify patterns.
In devising a plan and carrying it out, students adopt diverse strategies, including using pen and paper, and logical or creative thinking. Additionally, they decompose problems, take simple steps, work backwards from the problem question, and narrow down potential answers by process of~elimination. 

Beyond what students reported, our observations and an informal review of a sample of submissions provide some indication of how these strategies translated into practice. While we cannot be certain that students consistently applied what they wrote, in many cases they solved the tasks competently, and their plans and solutions clearly described how they did so.
When both a plan and a solution were present, they were generally logically aligned.
Yet, about 16\% of the students reported not having a clear strategy. %

In summary, by building on prior work emphasizing the relationship between CT and problem solving \cite{wuIdentificationProblemSolvingTechniques2024, iste_csta_ct_2011} and extending studies assessing the use of CT skills \cite{baveraComputationalThinkingSkills2020, dejongUseComputationalThinking2024} with a more in-depth qualitative analysis, we contribute new insights on the use of CT skills among higher-education students. 
They applied CT skills when solving Bebras tasks and, notably, they performed better when explicitly describing their plans and solution steps. 
However, some students reported experiencing difficulties across multiple problem-solving phases, such as planning, possibly because they were not used to applying problem-solving strategies systematically. 
These findings suggest that structured problem-solving practice could help beginner students, particularly those facing difficulties, develop CT skills and more deliberate problem-solving strategies.
Finally, we also demonstrate that Bebras tasks can be well-suited to challenge such students to apply CT skills in problem-solving scenarios, inviting research on exploring further uses of Bebras tasks in higher-education CT interventions and assessment.

\section{Limitations and Threats to Validity}
\label{sec:limitations}

This study has several limitations.
The collected data came from non-CS major students with diverse backgrounds who voluntarily participated. They were all enrolled in a single course at a particular university, thus our findings cannot be generalized to all undergraduate students in introductory programming.
As our collected demographics information was limited and our quantitative analysis was exploratory, we cannot make strong claims based on student characteristics.
Furthermore, our method has inherent limitations regarding which skills and parts of the problem-solving process can be observed.
We administered only five Bebras tasks due to constraints on the workload for assignments during the course. The students were asked to describe their solutions in written form, which requires more effort than selecting the correct answer in a multiple-choice format.
When selecting tasks, we prioritized problem variety and a suitable difficulty curve, meaning that the chosen Bebras tasks could not cover all CT skills equally, which is a recognized challenge with Bebras tasks \cite{dagieneDevelopingTwoDimensionalCategorization2017, izuExploringBebrasTasks2017}.
Additionally, we may have missed expressions of CT skills or other cognitive processes because we have no access to students' verbalizations while they were solving the tasks. 
Students may also have changed the traces of how they solved a problem as they had a week to submit, unsupervised. 
These factors could have influenced the distribution of observed CT skill expressions. 
The joint analysis of think-aloud transcripts and written artifacts could be employed to gather more data about the problem-solving process.

\section{Conclusion and Future Work}
\label{sec:conclusion}

This study investigates how higher-education students in an introductory programming course apply CT skills when solving computational problems and examines their difficulties and strategies.
We collected plans and solutions to Bebras tasks from 58 students, and analyzed the data using mixed methods.
RQ1 investigates how students apply CT skills when solving the tasks.
The most prevalent CT skills were algorithmic thinking, abstraction, and decomposition. The task type and the student's approach to structuring and describing the solution influenced which CT skills were observable and to what extent. 
Additionally, correct answers were more likely to contain plans and CT skills compared to incorrect ones.
RQ2 studies the difficulties encountered by students. We found that multiple aspects of the tasks and of the problem-solving process, including planning, proved challenging for students. 
RQ3 investigates the strategies students use to solve tasks.
Students employed and reported multiple meta-cognitive strategies.
Not all students could articulate their problem-solving strategies and processes, and reach the correct answer.

Together, these findings suggest that Bebras tasks are suitable for eliciting CT skills among introductory programming students.
They also provide a starting point to further investigate whether deliberate, scaffolded practice with problems such as Bebras tasks can support the development of CT skills and problem-solving strategies.
Future work could inquire into how background, motivation, and self-efficacy can affect students' use of CT skills and problem-solving. 
Next, we plan to explore approaches and tools to foster CT skills among introductory programming students by providing feedback during the problem-solving process.

\label{sec:acknowledgments}

\balance
\bibliographystyle{ACM-Reference-Format}
\bibliography{main.bib}


\begin{thebibliography}{65}

%%% the Web address.

\ifx \showCODEN    \undefined \def \showCODEN     #1{\unskip}     \fi
\ifx \showISBNx    \undefined \def \showISBNx     #1{\unskip}     \fi
\ifx \showISBNxiii \undefined \def \showISBNxiii  #1{\unskip}     \fi
\ifx \showISSN     \undefined \def \showISSN      #1{\unskip}     \fi
\ifx \showLCCN     \undefined \def \showLCCN      #1{\unskip}     \fi
\ifx \shownote     \undefined \def \shownote      #1{#1}          \fi
\ifx \showarticletitle \undefined \def \showarticletitle #1{#1}   \fi
\ifx \showURL      \undefined \def \showURL       {\relax}        \fi
% invisible to TeX
\providecommand\bibfield[2]{#2}
\providecommand\bibinfo[2]{#2}
\providecommand\natexlab[1]{#1}
\providecommand\showeprint[2][]{arXiv:#2}

\bibitem[Aho(2011)]%
        {ahoUbiquitySymposiumComputation2011}
\bibfield{author}{\bibinfo{person}{Alfred~V. Aho}.}
  \bibinfo{year}{2011}\natexlab{}.
\newblock \showarticletitle{Ubiquity symposium: Computation and Computational
  Thinking}.
\newblock \bibinfo{journal}{\emph{Ubiquity}} \bibinfo{volume}{2011},
  \bibinfo{number}{January}, Article \bibinfo{articleno}{1}
  (\bibinfo{date}{Jan.} \bibinfo{year}{2011}), \bibinfo{numpages}{8}~pages.
\newblock
\href{https://doi.org/10.1145/1922681.1922682}{doi:\nolinkurl{10.1145/1922681.1922682}}


\bibitem[Araujo et~al\mbox{.}(2019)]%
        {araujoHowManyAbilities2019}
\bibfield{author}{\bibinfo{person}{Ana Liz Souto~O. Araujo},
  \bibinfo{person}{Wilkerson~L. Andrade}, \bibinfo{person}{Dalton D.~Serey
  Guerrero}, {and} \bibinfo{person}{Monilly Ramos~Araujo Melo}.}
  \bibinfo{year}{2019}\natexlab{}.
\newblock \showarticletitle{How {{Many Abilities Can We Measure}} in
  {{Computational Thinking}}? {{A Study}} on {{Bebras Challenge}}}. In
  \bibinfo{booktitle}{\emph{Proceedings of the 50th {{ACM Technical Symposium}}
  on {{Computer Science Education}}}} \emph{(\bibinfo{series}{{{SIGCSE}}
  '19})}. \bibinfo{publisher}{Association for Computing Machinery},
  \bibinfo{address}{New York, NY, USA}, \bibinfo{pages}{545--551}.
\newblock
\showISBNx{978-1-4503-5890-3}
\href{https://doi.org/10.1145/3287324.3287405}{doi:\nolinkurl{10.1145/3287324.3287405}}


\bibitem[Atmatzidou and Demetriadis(2016)]%
        {atmatzidouAdvancingStudentsComputational2016}
\bibfield{author}{\bibinfo{person}{Soumela Atmatzidou} {and}
  \bibinfo{person}{Stavros Demetriadis}.} \bibinfo{year}{2016}\natexlab{}.
\newblock \showarticletitle{Advancing Students' Computational Thinking Skills
  through Educational Robotics: {{A}} Study on Age and Gender Relevant
  Differences}.
\newblock \bibinfo{journal}{\emph{Robotics and Autonomous Systems}}
  \bibinfo{volume}{75} (\bibinfo{date}{Jan.} \bibinfo{year}{2016}),
  \bibinfo{pages}{661--670}.
\newblock
\showISSN{0921-8890}
\href{https://doi.org/10.1016/j.robot.2015.10.008}{doi:\nolinkurl{10.1016/j.robot.2015.10.008}}


\bibitem[Barendsen et~al\mbox{.}(2015)]%
        {barendsenConceptsK9Computer2015}
\bibfield{author}{\bibinfo{person}{Erik Barendsen}, \bibinfo{person}{Linda
  Mannila}, \bibinfo{person}{Barbara Demo}, \bibinfo{person}{Nata\v{s}a
  Grgurina}, \bibinfo{person}{Cruz Izu}, \bibinfo{person}{Claudio Mirolo},
  \bibinfo{person}{Sue Sentance}, \bibinfo{person}{Amber Settle}, {and}
  \bibinfo{person}{Gabrielundefined Stupurienundefined}.}
  \bibinfo{year}{2015}\natexlab{}.
\newblock \showarticletitle{Concepts in K-9 Computer Science Education}. In
  \bibinfo{booktitle}{\emph{Proceedings of the 2015 ITiCSE on Working Group
  Reports}} (Vilnius, Lithuania) \emph{(\bibinfo{series}{ITICSE-WGR '15})}.
  \bibinfo{publisher}{Association for Computing Machinery},
  \bibinfo{address}{New York, NY, USA}, \bibinfo{pages}{85–116}.
\newblock
\showISBNx{9781450341462}
\href{https://doi.org/10.1145/2858796.2858800}{doi:\nolinkurl{10.1145/2858796.2858800}}


\bibitem[Bavera et~al\mbox{.}(2020)]%
        {baveraComputationalThinkingSkills2020}
\bibfield{author}{\bibinfo{person}{Francisco Bavera}, \bibinfo{person}{Teresa
  Quintero}, \bibinfo{person}{Marcela Daniele}, {and} \bibinfo{person}{Flavia
  Buffarini}.} \bibinfo{year}{2020}\natexlab{}.
\newblock \showarticletitle{Computational {{Thinking Skills}} in {{Primary
  Teachers}}: {{Evaluation Using Bebras}}}. In
  \bibinfo{booktitle}{\emph{Computer {{Science}} – {{CACIC}} 2019}} (Cham),
  \bibfield{editor}{\bibinfo{person}{Patricia Pesado} {and}
  \bibinfo{person}{Marcelo Arroyo}} (Eds.). \bibinfo{publisher}{Springer
  International Publishing}, \bibinfo{address}{Río Cuarto, Argentina},
  \bibinfo{pages}{405--415}.
\newblock
\showISBNx{978-3-030-48325-8}
\href{https://doi.org/10.1007/978-3-030-48325-8_26}{doi:\nolinkurl{10.1007/978-3-030-48325-8_26}}


\bibitem[Berkaliev et~al\mbox{.}(2014)]%
        {berkalievInitiatingProgrammaticAssessment2014}
\bibfield{author}{\bibinfo{person}{Zaur Berkaliev}, \bibinfo{person}{Shavila
  Devi}, \bibinfo{person}{Gregory~E. Fasshauer}, \bibinfo{person}{Fred~J.
  Hickernell}, \bibinfo{person}{Ozgul Kartal}, \bibinfo{person}{Xiaofan Li},
  \bibinfo{person}{Patrick McCray}, \bibinfo{person}{Stephanie Whitney}, {and}
  \bibinfo{person}{Judith~S. Zawojewski}.} \bibinfo{year}{2014}\natexlab{}.
\newblock \showarticletitle{Initiating a {{Programmatic Assessment Report}}}.
\newblock \bibinfo{journal}{\emph{PRIMUS}} \bibinfo{volume}{24},
  \bibinfo{number}{5} (\bibinfo{date}{May} \bibinfo{year}{2014}),
  \bibinfo{pages}{403--420}.
\newblock
\showISSN{1051-1970}
\href{https://doi.org/10.1080/10511970.2014.893939}{doi:\nolinkurl{10.1080/10511970.2014.893939}}


\bibitem[Berland and Lee(2011)]%
        {berlandCollaborativeStrategicBoard2011}
\bibfield{author}{\bibinfo{person}{Matthew Berland} {and}
  \bibinfo{person}{Victor~R. Lee}.} \bibinfo{year}{2011}\natexlab{}.
\newblock \showarticletitle{Collaborative {{Strategic Board Games}} as a
  {{Site}} for {{Distributed Computational Thinking}}:}.
\newblock \bibinfo{journal}{\emph{International Journal of Game-Based
  Learning}} \bibinfo{volume}{1}, \bibinfo{number}{2} (\bibinfo{date}{April}
  \bibinfo{year}{2011}), \bibinfo{pages}{65--81}.
\newblock
\showISSN{2155-6849, 2155-6857}
\href{https://doi.org/10.4018/ijgbl.2011040105}{doi:\nolinkurl{10.4018/ijgbl.2011040105}}


\bibitem[Bonner et~al\mbox{.}(2021)]%
        {bonnerFormativeAssessmentComputational2021}
\bibfield{author}{\bibinfo{person}{Sarah Bonner}, \bibinfo{person}{Peggy Chen},
  \bibinfo{person}{Kristi Jones}, {and} \bibinfo{person}{Brandon Milonovich}.}
  \bibinfo{year}{2021}\natexlab{}.
\newblock \showarticletitle{Formative {{Assessment}} of {{Computational
  Thinking}}: {{Cognitive}} and {{Metacognitive Processes}}}.
\newblock \bibinfo{journal}{\emph{Applied Measurement in Education}}
  \bibinfo{volume}{34}, \bibinfo{number}{1} (\bibinfo{date}{Jan.}
  \bibinfo{year}{2021}), \bibinfo{pages}{27--45}.
\newblock
\showISSN{0895-7347}
\href{https://doi.org/10.1080/08957347.2020.1835912}{doi:\nolinkurl{10.1080/08957347.2020.1835912}}


\bibitem[Braun and Clarke(2006)]%
        {braunUsingThematicAnalysis2006}
\bibfield{author}{\bibinfo{person}{Virginia Braun} {and}
  \bibinfo{person}{Victoria Clarke}.} \bibinfo{year}{2006}\natexlab{}.
\newblock \showarticletitle{Using Thematic Analysis in Psychology}.
\newblock \bibinfo{journal}{\emph{Qualitative Research in Psychology}}
  \bibinfo{volume}{3}, \bibinfo{number}{2} (\bibinfo{date}{Jan.}
  \bibinfo{year}{2006}), \bibinfo{pages}{77--101}.
\newblock
\showISSN{1478-0887}
\href{https://doi.org/10.1191/1478088706qp063oa}{doi:\nolinkurl{10.1191/1478088706qp063oa}}


\bibitem[Brennan and Resnick(2012)]%
        {brennanNewFrameworksStudying2012}
\bibfield{author}{\bibinfo{person}{Karen Brennan} {and}
  \bibinfo{person}{Mitchel Resnick}.} \bibinfo{year}{2012}\natexlab{}.
\newblock \showarticletitle{New frameworks for studying and assessing the
  development of computational thinking}. In
  \bibinfo{booktitle}{\emph{Proceedings of the 2012 annual meeting of the
  American educational research association}}, Vol.~\bibinfo{volume}{1}.
  \bibinfo{publisher}{AERA}, \bibinfo{address}{Vancouver, Canada},
  \bibinfo{pages}{25}.
\newblock


\bibitem[Corrales~{\'A}lvarez et~al\mbox{.}(2025)]%
        {corralesalvarezComputationalThinkingUniversity2025}
\bibfield{author}{\bibinfo{person}{Milena Corrales~{\'A}lvarez},
  \bibinfo{person}{Angela Mu{\~n}oz~Mu{\~n}oz}, {and} \bibinfo{person}{Sergio
  Cardona}.} \bibinfo{year}{2025}\natexlab{}.
\newblock \showarticletitle{Computational {{Thinking}} in the {{University
  Context}}: {{A Literature Review}} of {{Assessment Instruments}}}.
\newblock \bibinfo{journal}{\emph{TecnoL\'ogicas}}  \bibinfo{volume}{28}
  (\bibinfo{date}{Sept.} \bibinfo{year}{2025}), \bibinfo{pages}{e3394}.
\newblock
\href{https://doi.org/10.22430/22565337.3394}{doi:\nolinkurl{10.22430/22565337.3394}}


\bibitem[Czerkawski and Lyman(2015)]%
        {czerkawskiExploringIssuesComputational2015}
\bibfield{author}{\bibinfo{person}{Betul~C. Czerkawski} {and}
  \bibinfo{person}{Eugene~W. Lyman}.} \bibinfo{year}{2015}\natexlab{}.
\newblock \showarticletitle{Exploring {{Issues About Computational Thinking}}
  in {{Higher Education}}}.
\newblock \bibinfo{journal}{\emph{TechTrends}} \bibinfo{volume}{59},
  \bibinfo{number}{2} (\bibinfo{date}{March} \bibinfo{year}{2015}),
  \bibinfo{pages}{57--65}.
\newblock
\showISSN{1559-7075}
\href{https://doi.org/10.1007/s11528-015-0840-3}{doi:\nolinkurl{10.1007/s11528-015-0840-3}}


\bibitem[Dagiene and Dolgopolovas(2022)]%
        {dagieneShortTasksScaffolding2022}
\bibfield{author}{\bibinfo{person}{Valentina Dagiene} {and}
  \bibinfo{person}{Vladimiras Dolgopolovas}.} \bibinfo{year}{2022}\natexlab{}.
\newblock \showarticletitle{Short {{Tasks}} for {{Scaffolding Computational
  Thinking}} by the {{Global Bebras Challenge}}}.
\newblock \bibinfo{journal}{\emph{Mathematics}} \bibinfo{volume}{10},
  \bibinfo{number}{17} (\bibinfo{date}{Jan.} \bibinfo{year}{2022}),
  \bibinfo{pages}{3194}.
\newblock
\showISSN{2227-7390}
\href{https://doi.org/10.3390/math10173194}{doi:\nolinkurl{10.3390/math10173194}}


\bibitem[Dagien{\.e} and Sentance(2016)]%
        {dagieneItsComputationalThinking2016b}
\bibfield{author}{\bibinfo{person}{Valentina Dagien{\.e}} {and}
  \bibinfo{person}{Sue Sentance}.} \bibinfo{year}{2016}\natexlab{}.
\newblock \showarticletitle{It's {{Computational Thinking}}! {{Bebras Tasks}}
  in the {{Curriculum}}}. In \bibinfo{booktitle}{\emph{Informatics in
  {{Schools}}: {{Improvement}} of {{Informatics Knowledge}} and
  {{Perception}}}}, \bibfield{editor}{\bibinfo{person}{Andrej Brodnik} {and}
  \bibinfo{person}{Fran{\c c}oise Tort}} (Eds.). \bibinfo{publisher}{Springer
  International Publishing}, \bibinfo{address}{Cham}, \bibinfo{pages}{28--39}.
\newblock
\showISBNx{978-3-319-46747-4}
\href{https://doi.org/10.1007/978-3-319-46747-4_3}{doi:\nolinkurl{10.1007/978-3-319-46747-4_3}}


\bibitem[Dagien{\.e} et~al\mbox{.}(2017)]%
        {dagieneDevelopingTwoDimensionalCategorization2017}
\bibfield{author}{\bibinfo{person}{Valentina Dagien{\.e}}, \bibinfo{person}{Sue
  Sentance}, {and} \bibinfo{person}{Gabriel{\.e} Stupurien{\.e}}.}
  \bibinfo{year}{2017}\natexlab{}.
\newblock \showarticletitle{Developing a {{Two-Dimensional Categorization
  System}} for {{Educational Tasks}} in {{Informatics}}}.
\newblock \bibinfo{journal}{\emph{Informatica}} \bibinfo{volume}{28},
  \bibinfo{number}{1} (\bibinfo{date}{Feb.} \bibinfo{year}{2017}),
  \bibinfo{pages}{23--44}.
\newblock
\showISSN{0868-4952}
\href{https://doi.org/10.3233/INF-2017-1127}{doi:\nolinkurl{10.3233/INF-2017-1127}}


\bibitem[{de Jong} and Jeuring(2020)]%
        {dejongComputationalThinkingInterventions2020}
\bibfield{author}{\bibinfo{person}{Imke {de Jong}} {and} \bibinfo{person}{Johan
  Jeuring}.} \bibinfo{year}{2020}\natexlab{}.
\newblock \showarticletitle{Computational {{Thinking Interventions}} in
  {{Higher Education}}: {{A Scoping Literature Review}} of {{Interventions
  Used}} to {{Teach Computational Thinking}}}. In
  \bibinfo{booktitle}{\emph{Proceedings of the 20th {{Koli Calling
  International Conference}} on {{Computing Education Research}}}}
  \emph{(\bibinfo{series}{Koli {{Calling}} '20})}.
  \bibinfo{publisher}{Association for Computing Machinery},
  \bibinfo{address}{New York, NY, USA}, \bibinfo{pages}{1--10}.
\newblock
\showISBNx{978-1-4503-8921-1}
\href{https://doi.org/10.1145/3428029.3428055}{doi:\nolinkurl{10.1145/3428029.3428055}}


\bibitem[{de Jong} et~al\mbox{.}(2024)]%
        {dejongUseComputationalThinking2024}
\bibfield{author}{\bibinfo{person}{Imke {de Jong}}, \bibinfo{person}{Bo
  Sichterman}, {and} \bibinfo{person}{Johan Jeuring}.}
  \bibinfo{year}{2024}\natexlab{}.
\newblock \showarticletitle{Use of {{Computational Thinking Skills}} When
  Solving {{Bebras Tasks}}: A {{Think-aloud Study}}}. In
  \bibinfo{booktitle}{\emph{Proceedings of the 24th {{Koli Calling
  International Conference}} on {{Computing Education Research}}}}
  \emph{(\bibinfo{series}{Koli {{Calling}} '24})}.
  \bibinfo{publisher}{Association for Computing Machinery},
  \bibinfo{address}{New York, NY, USA}, \bibinfo{pages}{1--11}.
\newblock
\showISBNx{979-8-4007-1038-4}
\href{https://doi.org/10.1145/3699538.3699543}{doi:\nolinkurl{10.1145/3699538.3699543}}


\bibitem[{diSessa}(2018)]%
        {disessaComputationalLiteracyBig2018}
\bibfield{author}{\bibinfo{person}{Andrea~A. {diSessa}}.}
  \bibinfo{year}{2018}\natexlab{}.
\newblock \showarticletitle{Computational {{Literacy}} and “{{The Big
  Picture}}” {{Concerning Computers}} in {{Mathematics Education}}}.
\newblock \bibinfo{journal}{\emph{Mathematical Thinking and Learning}}
  \bibinfo{volume}{20}, \bibinfo{number}{1} (\bibinfo{year}{2018}),
  \bibinfo{pages}{3--31}.
\newblock
\showISSN{1098-6065}
\href{https://doi.org/10.1080/10986065.2018.1403544}{doi:\nolinkurl{10.1080/10986065.2018.1403544}}


\bibitem[Febrian et~al\mbox{.}(2018)]%
        {febrianDoesEveryoneUse2018}
\bibfield{author}{\bibinfo{person}{Andreas Febrian}, \bibinfo{person}{Oenardi
  Lawanto}, \bibinfo{person}{Kamyn {Peterson-Rucker}}, \bibinfo{person}{Alia
  Melvin}, {and} \bibinfo{person}{Shane~E. Guymon}.}
  \bibinfo{year}{2018}\natexlab{}.
\newblock \showarticletitle{Does {{Everyone Use Computational Thinking}}?: {{A
  Case Study}} of {{Art}} and {{Computer Science Majors}}}. In
  \bibinfo{booktitle}{\emph{2018 {{ASEE Annual Conference}} \&
  {{Exposition}}}}.
\newblock


\bibitem[Graham et~al\mbox{.}(2012)]%
        {grahamMeasuringPromotingInterrater}
\bibfield{author}{\bibinfo{person}{Matthew Graham}, \bibinfo{person}{Anthony
  Milanowski}, {and} \bibinfo{person}{Jackson Miller}.}
  \bibinfo{year}{2012}\natexlab{}.
\newblock \showarticletitle{Measuring and Promoting Inter-Rater Agreement of
  Teacher and Principal Performance Ratings.}
\newblock \bibinfo{journal}{\emph{Online Submission}} (\bibinfo{year}{2012}).
\newblock


\bibitem[Grigg and Benson(2012)]%
        {griggEffectsStudentStrategies2012}
\bibfield{author}{\bibinfo{person}{Sarah~Jane Grigg} {and}
  \bibinfo{person}{Lisa Benson}.} \bibinfo{year}{2012}\natexlab{}.
\newblock \showarticletitle{Effects of {{Student Strategies}} on {{Successful
  Problem Solving}}}. In \bibinfo{booktitle}{\emph{2012 {{ASEE Annual
  Conference}} \& {{Exposition}}}}. \bibinfo{pages}{25.508.1--25.508.13}.
\newblock
\showISSN{2153-5965}
\href{https://doi.org/10.18260/1-2--21266}{doi:\nolinkurl{10.18260/1-2--21266}}


\bibitem[Grover and Pea(2013)]%
        {groverComputationalThinkingK122013a}
\bibfield{author}{\bibinfo{person}{Shuchi Grover} {and} \bibinfo{person}{Roy
  Pea}.} \bibinfo{year}{2013}\natexlab{}.
\newblock \showarticletitle{Computational {{Thinking}} in {{K}}–12: {{A
  Review}} of the {{State}} of the {{Field}}}.
\newblock \bibinfo{journal}{\emph{Educational Researcher}}
  \bibinfo{volume}{42}, \bibinfo{number}{1} (\bibinfo{year}{2013}),
  \bibinfo{pages}{38--43}.
\newblock
\showISSN{0013-189X}
\href{https://doi.org/10.3102/0013189X12463051}{doi:\nolinkurl{10.3102/0013189X12463051}}


\bibitem[Hsu et~al\mbox{.}(2018)]%
        {hsuHowLearnHow2018a}
\bibfield{author}{\bibinfo{person}{Ting-Chia Hsu}, \bibinfo{person}{Shao-Chen
  Chang}, {and} \bibinfo{person}{Yu-Ting Hung}.}
  \bibinfo{year}{2018}\natexlab{}.
\newblock \showarticletitle{How to Learn and How to Teach Computational
  Thinking: {{Suggestions}} Based on a Review of the Literature}.
\newblock \bibinfo{journal}{\emph{Computers \& Education}}
  \bibinfo{volume}{126} (\bibinfo{date}{Nov.} \bibinfo{year}{2018}),
  \bibinfo{pages}{296--310}.
\newblock
\showISSN{0360-1315}
\href{https://doi.org/10.1016/j.compedu.2018.07.004}{doi:\nolinkurl{10.1016/j.compedu.2018.07.004}}


\bibitem[{International Society for Technology in Education} and {Computer
  Science Teachers Association}(2011)]%
        {iste_csta_ct_2011}
\bibfield{author}{\bibinfo{person}{{International Society for Technology in
  Education}} {and} \bibinfo{person}{{Computer Science Teachers Association}}.}
  \bibinfo{year}{2011}\natexlab{}.
\newblock \bibinfo{booktitle}{\emph{Computational {{Thinking}} in {{K}}--12
  {{Education}}: {{Teacher Resources}}} (\bibinfo{edition}{second} ed.)}.
\newblock {ISTE and CSTA}.
\newblock
\urldef\tempurl%
\url{https://cdn.iste.org/www-root/2020-10/ISTE_CT_Teacher_Resources_2ed.pdf}
\showURL{%
\tempurl}


\bibitem[Izu et~al\mbox{.}(2017)]%
        {izuExploringBebrasTasks2017}
\bibfield{author}{\bibinfo{person}{Cruz Izu}, \bibinfo{person}{Claudio Mirolo},
  \bibinfo{person}{Amber Settle}, \bibinfo{person}{Linda Mannila}, {and}
  \bibinfo{person}{Gabriele Stupuriene}.} \bibinfo{year}{2017}\natexlab{}.
\newblock \showarticletitle{Exploring Bebras Tasks Content and Performance:
  {{A}} Multinational Study}.
\newblock \bibinfo{journal}{\emph{Informatics in Education}}
  \bibinfo{volume}{16}, \bibinfo{number}{1} (\bibinfo{year}{2017}),
  \bibinfo{pages}{39--59}.
\newblock
\showISSN{1648-5831}
\href{https://doi.org/10.15388/infedu.2017.03}{doi:\nolinkurl{10.15388/infedu.2017.03}}


\bibitem[Kalelioglu et~al\mbox{.}(2016)]%
        {kaleliogluFrameworkComputationalThinking2016}
\bibfield{author}{\bibinfo{person}{Filiz Kalelioglu}, \bibinfo{person}{Yasemin
  Gulbahar}, {and} \bibinfo{person}{Volkan Kukul}.}
  \bibinfo{year}{2016}\natexlab{}.
\newblock \showarticletitle{A Framework for Computational Thinking Based on a
  Systematic Research Review}.
\newblock \bibinfo{journal}{\emph{Baltic Journal of Modern Computing}}
  \bibinfo{volume}{4} (\bibinfo{date}{05} \bibinfo{year}{2016}),
  \bibinfo{pages}{583--596}.
\newblock


\bibitem[{Kastner-Hauler} et~al\mbox{.}(2024)]%
        {kastner-haulerLearningEnvironmentPromote2024}
\bibfield{author}{\bibinfo{person}{Oliver {Kastner-Hauler}},
  \bibinfo{person}{Karin Tengler}, \bibinfo{person}{Barbara Sabitzer}, {and}
  \bibinfo{person}{Zsolt Lavicza}.} \bibinfo{year}{2024}\natexlab{}.
\newblock \showarticletitle{A {{Learning Environment}} to~{{Promote}}
  the~{{Computational Thinker}}: {{A Bebras Perspective Evaluation}}}. In
  \bibinfo{booktitle}{\emph{Informatics in {{Schools}}. {{Innovative
  Approaches}} to {{Computer Science Teaching}} and {{Learning}}}},
  \bibfield{editor}{\bibinfo{person}{Zsuzsa Pluh{\'a}r} {and}
  \bibinfo{person}{Bence Ga{\'a}l}} (Eds.). \bibinfo{publisher}{Springer Nature
  Switzerland}, \bibinfo{address}{Cham}, \bibinfo{pages}{85--98}.
\newblock
\showISBNx{978-3-031-73474-8}
\href{https://doi.org/10.1007/978-3-031-73474-8_7}{doi:\nolinkurl{10.1007/978-3-031-73474-8_7}}


\bibitem[Li et~al\mbox{.}(2020)]%
        {liComputationalThinkingMore2020}
\bibfield{author}{\bibinfo{person}{Yeping Li}, \bibinfo{person}{Alan~H.
  Schoenfeld}, \bibinfo{person}{Andrea~A. {diSessa}},
  \bibinfo{person}{Arthur~C. Graesser}, \bibinfo{person}{Lisa~C. Benson},
  \bibinfo{person}{Lyn~D. English}, {and} \bibinfo{person}{Richard~A. Duschl}.}
  \bibinfo{year}{2020}\natexlab{}.
\newblock \showarticletitle{Computational {{Thinking Is More}} about
  {{Thinking}} than {{Computing}}}.
\newblock \bibinfo{journal}{\emph{Journal for STEM Education Research}}
  \bibinfo{volume}{3}, \bibinfo{number}{1} (\bibinfo{year}{2020}),
  \bibinfo{pages}{1--18}.
\newblock
\showISSN{2520-8713}
\href{https://doi.org/10.1007/s41979-020-00030-2}{doi:\nolinkurl{10.1007/s41979-020-00030-2}}


\bibitem[Linn(1985)]%
        {linnCognitiveConsequencesProgramming1985}
\bibfield{author}{\bibinfo{person}{Marcia~C. Linn}.}
  \bibinfo{year}{1985}\natexlab{}.
\newblock \showarticletitle{The {{Cognitive Consequences}} of {{Programming
  Instruction}} in {{Classrooms}}}.
\newblock \bibinfo{journal}{\emph{Educational Researcher}}
  \bibinfo{volume}{14}, \bibinfo{number}{5} (\bibinfo{year}{1985}),
  \bibinfo{pages}{14--29}.
\newblock
\showISSN{0013-189X}
\showeprint[jstor]{1174202}
\href{https://doi.org/10.2307/1174202}{doi:\nolinkurl{10.2307/1174202}}


\bibitem[Lockwood and Mooney(2018)]%
        {lockwoodDevelopingComputationalThinking2018}
\bibfield{author}{\bibinfo{person}{James Lockwood} {and} \bibinfo{person}{Aidan
  Mooney}.} \bibinfo{year}{2018}\natexlab{}.
\newblock \showarticletitle{Developing a Computational Thinking Test using
  Bebras problems}. In \bibinfo{booktitle}{\emph{CC-TEL 2018 and TACKLE 2018
  Workshops}}.
\newblock
\urldef\tempurl%
\url{https://mural.maynoothuniversity.ie/id/eprint/10316/}
\showURL{%
\tempurl}
\newblock
\shownote{In: A. Piotrkowicz, R. Dent-Spargo, S. Dennerlein, I. Koren, P.
  Antoniou, P. Bailey, T. Treasure-Jones, I. Fronza, C. Pahl (eds.): Joint
  Proceedings of the CC-TEL 2018 and TACKLE 2018 Workshops, co-located with
  13th European Conference on Technology Enhanced Learning (EC-TEL 2018),
  03-09-2018, published at http://ceur-ws.org}.


\bibitem[Lu et~al\mbox{.}(2022)]%
        {luScopingReviewComputational2022}
\bibfield{author}{\bibinfo{person}{Chang Lu}, \bibinfo{person}{Rob Macdonald},
  \bibinfo{person}{Bryce Odell}, \bibinfo{person}{Vasyl Kokhan},
  \bibinfo{person}{Carrie Demmans~Epp}, {and} \bibinfo{person}{Maria
  Cutumisu}.} \bibinfo{year}{2022}\natexlab{}.
\newblock \showarticletitle{A Scoping Review of Computational Thinking
  Assessments in Higher Education}.
\newblock \bibinfo{journal}{\emph{Journal of Computing in Higher Education}}
  \bibinfo{volume}{34}, \bibinfo{number}{2} (\bibinfo{date}{Aug.}
  \bibinfo{year}{2022}), \bibinfo{pages}{416--461}.
\newblock
\showISSN{1867-1233}
\href{https://doi.org/10.1007/s12528-021-09305-y}{doi:\nolinkurl{10.1007/s12528-021-09305-y}}


\bibitem[Luo et~al\mbox{.}(2020)]%
        {luoUnderstandingStudentsComputational2020a}
\bibfield{author}{\bibinfo{person}{Feiya Luo}, \bibinfo{person}{Maya Israel},
  \bibinfo{person}{Ruohan Liu}, \bibinfo{person}{Wei Yan},
  \bibinfo{person}{Brian Gane}, {and} \bibinfo{person}{John Hampton}.}
  \bibinfo{year}{2020}\natexlab{}.
\newblock \showarticletitle{Understanding {{Students}}' {{Computational
  Thinking}} through {{Cognitive Interviews}}: {{A Learning Trajectory-based
  Analysis}}}. In \bibinfo{booktitle}{\emph{Proceedings of the 51st {{ACM
  Technical Symposium}} on {{Computer Science Education}}}}
  \emph{(\bibinfo{series}{{{SIGCSE}} '20})}. \bibinfo{publisher}{Association
  for Computing Machinery}, \bibinfo{address}{New York, NY, USA},
  \bibinfo{pages}{919--925}.
\newblock
\showISBNx{978-1-4503-6793-6}
\href{https://doi.org/10.1145/3328778.3366845}{doi:\nolinkurl{10.1145/3328778.3366845}}


\bibitem[Lyon and J.~Magana(2020)]%
        {lyonComputationalThinkingHigher2020}
\bibfield{author}{\bibinfo{person}{Joseph~A. Lyon} {and}
  \bibinfo{person}{Alejandra J.~Magana}.} \bibinfo{year}{2020}\natexlab{}.
\newblock \showarticletitle{Computational Thinking in Higher Education: {{A}}
  Review of the Literature}.
\newblock \bibinfo{journal}{\emph{Computer Applications in Engineering
  Education}} \bibinfo{volume}{28}, \bibinfo{number}{5} (\bibinfo{year}{2020}),
  \bibinfo{pages}{1174--1189}.
\newblock
\showISSN{1099-0542}
\href{https://doi.org/10.1002/cae.22295}{doi:\nolinkurl{10.1002/cae.22295}}


\bibitem[Lyon et~al\mbox{.}(2022)]%
        {lyonCharacterizingComputationalThinking2022}
\bibfield{author}{\bibinfo{person}{Joseph~A. Lyon},
  \bibinfo{person}{Alejandra~J. Magana}, {and} \bibinfo{person}{Ruth~A.
  Streveler}.} \bibinfo{year}{2022}\natexlab{}.
\newblock \showarticletitle{Characterizing {{Computational Thinking}} in the
  {{Context}} of {{Model-Planning Activities}}}.
\newblock \bibinfo{journal}{\emph{Modelling 2022}} \bibinfo{volume}{3},
  \bibinfo{number}{3} (\bibinfo{year}{2022}), \bibinfo{pages}{344--358}.
\newblock
\showISSN{2673-3951}
\href{https://doi.org/10.3390/MODELLING3030022}{doi:\nolinkurl{10.3390/MODELLING3030022}}


\bibitem[Ma et~al\mbox{.}(2025)]%
        {maDBoxScaffoldingAlgorithmic2025}
\bibfield{author}{\bibinfo{person}{Shuai Ma}, \bibinfo{person}{Junling Wang},
  \bibinfo{person}{Yuanhao Zhang}, \bibinfo{person}{Xiaojuan Ma}, {and}
  \bibinfo{person}{April~Yi Wang}.} \bibinfo{year}{2025}\natexlab{}.
\newblock \showarticletitle{DBox: Scaffolding Algorithmic Programming Learning
  through Learner-LLM Co-Decomposition}. In
  \bibinfo{booktitle}{\emph{Proceedings of the 2025 CHI Conference on Human
  Factors in Computing Systems}} \emph{(\bibinfo{series}{CHI '25})}.
  \bibinfo{publisher}{Association for Computing Machinery},
  \bibinfo{address}{New York, NY, USA}, Article \bibinfo{articleno}{585},
  \bibinfo{numpages}{20}~pages.
\newblock
\showISBNx{9798400713941}
\href{https://doi.org/10.1145/3706598.3713748}{doi:\nolinkurl{10.1145/3706598.3713748}}


\bibitem[Margulieux et~al\mbox{.}(2019)]%
        {margulieuxReviewMeasurementsUsed2019}
\bibfield{author}{\bibinfo{person}{Lauren Margulieux},
  \bibinfo{person}{Tuba~Ayer Ketenci}, {and} \bibinfo{person}{Adrienne
  Decker}.} \bibinfo{year}{2019}\natexlab{}.
\newblock \showarticletitle{Review of Measurements Used in Computing Education
  Research and Suggestions for Increasing Standardization}.
\newblock \bibinfo{journal}{\emph{Computer Science Education}}
  \bibinfo{volume}{29}, \bibinfo{number}{1} (\bibinfo{date}{Jan.}
  \bibinfo{year}{2019}), \bibinfo{pages}{49--78}.
\newblock
\showISSN{0899-3408, 1744-5175}
\href{https://doi.org/10.1080/08993408.2018.1562145}{doi:\nolinkurl{10.1080/08993408.2018.1562145}}


\bibitem[Marwan et~al\mbox{.}(2024)]%
        {marwanExploringNovicesProblemSolving2024}
\bibfield{author}{\bibinfo{person}{Samiha Marwan}, \bibinfo{person}{Nicki
  Choquette}, \bibinfo{person}{Veronica Catet{\'e}}, {and}
  \bibinfo{person}{Briana B.~Morrison}.} \bibinfo{year}{2024}\natexlab{}.
\newblock \showarticletitle{Exploring {{Novices}}' {{Problem-Solving
  Strategies}} in {{Computing}} and {{Math Domains}}}. In
  \bibinfo{booktitle}{\emph{Proceedings of the 24th {{Koli Calling
  International Conference}} on {{Computing Education Research}}}}.
  \bibinfo{publisher}{ACM}, \bibinfo{address}{Koli Finland},
  \bibinfo{pages}{1--8}.
\newblock
\showISBNx{979-8-4007-1038-4}
\href{https://doi.org/10.1145/3699538.3699557}{doi:\nolinkurl{10.1145/3699538.3699557}}


\bibitem[Miller et~al\mbox{.}(2013)]%
        {millerImprovingLearningComputational2013}
\bibfield{author}{\bibinfo{person}{L.~Dee Miller}, \bibinfo{person}{Leen-Kiat
  Soh}, \bibinfo{person}{Vlad Chiriacescu}, \bibinfo{person}{Elizabeth
  Ingraham}, \bibinfo{person}{Duane~F. Shell}, \bibinfo{person}{Stephen
  Ramsay}, {and} \bibinfo{person}{Melissa~Patterson Hazley}.}
  \bibinfo{year}{2013}\natexlab{}.
\newblock \showarticletitle{Improving Learning of Computational Thinking Using
  Creative Thinking Exercises in {{CS-1}} Computer Science Courses}. In
  \bibinfo{booktitle}{\emph{2013 {{IEEE Frontiers}} in {{Education Conference}}
  ({{FIE}})}}. \bibinfo{publisher}{IEEE Computer Society},
  \bibinfo{address}{Los Alamitos, CA, USA}, \bibinfo{pages}{1426--1432}.
\newblock
\showISSN{2377-634X}
\href{https://doi.org/10.1109/FIE.2013.6685067}{doi:\nolinkurl{10.1109/FIE.2013.6685067}}


\bibitem[Mills et~al\mbox{.}(2024)]%
        {millsCodingComputationalThinking2024}
\bibfield{author}{\bibinfo{person}{Kathy~A. Mills}, \bibinfo{person}{Jen Cope},
  \bibinfo{person}{Laura Scholes}, {and} \bibinfo{person}{Luke Rowe}.}
  \bibinfo{year}{2024}\natexlab{}.
\newblock \showarticletitle{Coding and {{Computational Thinking Across}} the
  {{Curriculum}}: {{A Review}} of {{Educational Outcomes}}}.
\newblock \bibinfo{journal}{\emph{Review of Educational Research}}
  (\bibinfo{date}{April} \bibinfo{year}{2024}),
  \bibinfo{pages}{00346543241241327}.
\newblock
\showISSN{0034-6543}
\href{https://doi.org/10.3102/00346543241241327}{doi:\nolinkurl{10.3102/00346543241241327}}


\bibitem[Morrison et~al\mbox{.}(2014)]%
        {morrisonMeasuringCognitiveLoad2014}
\bibfield{author}{\bibinfo{person}{Briana~B. Morrison}, \bibinfo{person}{Brian
  Dorn}, {and} \bibinfo{person}{Mark Guzdial}.}
  \bibinfo{year}{2014}\natexlab{}.
\newblock \showarticletitle{Measuring Cognitive Load in Introductory {{CS}}:
  Adaptation of an Instrument}. In \bibinfo{booktitle}{\emph{Proceedings of the
  Tenth Annual Conference on {{International}} Computing Education Research}}.
  \bibinfo{publisher}{ACM}, \bibinfo{address}{Glasgow Scotland United Kingdom},
  \bibinfo{pages}{131--138}.
\newblock
\showISBNx{978-1-4503-2755-8}
\href{https://doi.org/10.1145/2632320.2632348}{doi:\nolinkurl{10.1145/2632320.2632348}}


\bibitem[Oliveira et~al\mbox{.}(2025)]%
        {oliveiraQuantifyingComputationalThinking2025}
\bibfield{author}{\bibinfo{person}{Ana Liz~Souto Oliveira},
  \bibinfo{person}{Wilkerson~L. Andrade}, \bibinfo{person}{Dalton Serey}, {and}
  \bibinfo{person}{Monilly Ramos~Araujo Melo}.}
  \bibinfo{year}{2025}\natexlab{}.
\newblock \showarticletitle{Quantifying {{Computational Thinking Skills}}: An
  {{Exploratory Study}} on {{Bebras Tasks}}}.
\newblock \bibinfo{journal}{\emph{Journal of the Brazilian Computer Society}}
  \bibinfo{volume}{31}, \bibinfo{number}{1} (\bibinfo{year}{2025}),
  \bibinfo{pages}{338--354}.
\newblock
Issue 1.
\showISSN{1678-4804}
\href{https://doi.org/10.5753/jbcs.2025.3893}{doi:\nolinkurl{10.5753/jbcs.2025.3893}}


\bibitem[Palts and Pedaste(2020)]%
        {paltsModelDevelopingComputational2020a}
\bibfield{author}{\bibinfo{person}{Tauno Palts} {and} \bibinfo{person}{Margus
  Pedaste}.} \bibinfo{year}{2020}\natexlab{}.
\newblock \showarticletitle{A {{Model}} for {{Developing Computational Thinking
  Skills}}}.
\newblock \bibinfo{journal}{\emph{Informatics in Education}}
  \bibinfo{volume}{19}, \bibinfo{number}{1} (\bibinfo{date}{March}
  \bibinfo{year}{2020}), \bibinfo{pages}{113--128}.
\newblock
\showISSN{1648-5831, 2335-8971}
\href{https://doi.org/10.15388/infedu.2020.06}{doi:\nolinkurl{10.15388/infedu.2020.06}}


\bibitem[Papert(1980)]%
        {papertMindstormsChildrenComputers1980}
\bibfield{author}{\bibinfo{person}{Seymour Papert}.}
  \bibinfo{year}{1980}\natexlab{}.
\newblock \bibinfo{booktitle}{\emph{Mindstorms: Children, Computers, and
  Powerful Ideas}}.
\newblock \bibinfo{publisher}{Basic Books, Inc.}, \bibinfo{address}{USA}.
\newblock
\showISBNx{978-0-465-04627-0}


\bibitem[Pea and Kurland(1984)]%
        {peaCognitiveEffectsLearning1984}
\bibfield{author}{\bibinfo{person}{Roy~D. Pea} {and} \bibinfo{person}{D.~Midian
  Kurland}.} \bibinfo{year}{1984}\natexlab{}.
\newblock \showarticletitle{On the Cognitive Effects of Learning Computer
  Programming}.
\newblock \bibinfo{journal}{\emph{New Ideas in Psychology}}
  \bibinfo{volume}{2}, \bibinfo{number}{2} (\bibinfo{date}{Jan.}
  \bibinfo{year}{1984}), \bibinfo{pages}{137--168}.
\newblock
\showISSN{0732-118X}
\href{https://doi.org/10.1016/0732-118X(84)90018-7}{doi:\nolinkurl{10.1016/0732-118X(84)90018-7}}


\bibitem[Polya(1945)]%
        {polyaHowSolveIt1945a}
\bibfield{author}{\bibinfo{person}{George Polya}.}
  \bibinfo{year}{1945}\natexlab{}.
\newblock \bibinfo{booktitle}{\emph{How to {{Solve It}}: {{A New Aspect}} of
  {{Mathematical Method}}}}.
\newblock \bibinfo{publisher}{Princeton University Press},
  \bibinfo{address}{Princeton, NJ}.
\newblock
\showISBNx{978-0-691-16407-6}
\showeprint[jstor]{j.ctvc773pk}
\href{https://doi.org/10.2307/j.ctvc773pk}{doi:\nolinkurl{10.2307/j.ctvc773pk}}


\bibitem[Polya(1981)]%
        {polyaMathematicalDiscoveryUnderstanding1981}
\bibfield{author}{\bibinfo{person}{George Polya}.}
  \bibinfo{year}{1981}\natexlab{}.
\newblock \bibinfo{booktitle}{\emph{Mathematical discovery : on understanding,
  learning, and teaching problem solving} (\bibinfo{edition}{combined ed.}
  ed.)}.
\newblock \bibinfo{publisher}{Wiley}.
\newblock
\urldef\tempurl%
\url{https://cir.nii.ac.jp/crid/1970586434849214760}
\showURL{%
\tempurl}


\bibitem[Pugalee(2004)]%
        {pugaleeComparisonVerbalWritten2004}
\bibfield{author}{\bibinfo{person}{David~K. Pugalee}.}
  \bibinfo{year}{2004}\natexlab{}.
\newblock \showarticletitle{A {{Comparison}} of {{Verbal}} and {{Written
  Descriptions}} of {{Students}}' {{Problem Solving Processes}}}.
\newblock \bibinfo{journal}{\emph{Educational Studies in Mathematics}}
  \bibinfo{volume}{55}, \bibinfo{number}{1} (\bibinfo{year}{2004}),
  \bibinfo{pages}{27--47}.
\newblock
\showISSN{1573-0816}
\href{https://doi.org/10.1023/B:EDUC.0000017666.11367.c7}{doi:\nolinkurl{10.1023/B:EDUC.0000017666.11367.c7}}


\bibitem[Raihan et~al\mbox{.}(2025)]%
        {raihanLargeLanguageModels2025}
\bibfield{author}{\bibinfo{person}{Nishat Raihan},
  \bibinfo{person}{Mohammed~Latif Siddiq}, \bibinfo{person}{Joanna~C.S.
  Santos}, {and} \bibinfo{person}{Marcos Zampieri}.}
  \bibinfo{year}{2025}\natexlab{}.
\newblock \showarticletitle{Large {{Language Models}} in {{Computer Science
  Education}}: {{A Systematic Literature Review}}}. In
  \bibinfo{booktitle}{\emph{Proceedings of the 56th {{ACM Technical Symposium}}
  on {{Computer Science Education V}}. 1}} \emph{(\bibinfo{series}{{{SIGCSETS}}
  2025})}. \bibinfo{publisher}{Association for Computing Machinery},
  \bibinfo{address}{New York, NY, USA}, \bibinfo{pages}{938--944}.
\newblock
\showISBNx{979-8-4007-0531-1}
\href{https://doi.org/10.1145/3641554.3701863}{doi:\nolinkurl{10.1145/3641554.3701863}}


\bibitem[Robins et~al\mbox{.}(2003)]%
        {robinsLearningTeachingProgramming2003}
\bibfield{author}{\bibinfo{person}{Anthony Robins}, \bibinfo{person}{Janet
  Rountree}, {and} \bibinfo{person}{Nathan Rountree}.}
  \bibinfo{year}{2003}\natexlab{}.
\newblock \showarticletitle{Learning and {{Teaching Programming}}: {{A Review}}
  and {{Discussion}}}.
\newblock \bibinfo{journal}{\emph{Computer Science Education}}
  \bibinfo{volume}{21}, \bibinfo{number}{1} (\bibinfo{year}{2003}),
  \bibinfo{pages}{137--172}.
\newblock
\showISSN{15497879}
\href{https://doi.org/10.1076/CSED.13.2.137.14200}{doi:\nolinkurl{10.1076/CSED.13.2.137.14200}}


\bibitem[Rom{\'a}n-Gonz{\'a}lez et~al\mbox{.}(2019)]%
        {romangonzalezCombiningAssessmentTools2019}
\bibfield{author}{\bibinfo{person}{Marcos Rom{\'a}n-Gonz{\'a}lez},
  \bibinfo{person}{Jes{\'u}s Moreno-Le{\'o}n}, {and} \bibinfo{person}{Gregorio
  Robles}.} \bibinfo{year}{2019}\natexlab{}.
\newblock \showarticletitle{Combining Assessment Tools for a Comprehensive
  Evaluation of Computational Thinking Interventions}.
\newblock In \bibinfo{booktitle}{\emph{Computational Thinking Education}},
  \bibfield{editor}{\bibinfo{person}{Siu-Cheung Kong} {and}
  \bibinfo{person}{Harold Abelson}} (Eds.). \bibinfo{publisher}{Springer
  Singapore}, \bibinfo{address}{Singapore}, \bibinfo{pages}{79--98}.
\newblock
\showISBNx{978-981-13-6528-7}
\href{https://doi.org/10.1007/978-981-13-6528-7_6}{doi:\nolinkurl{10.1007/978-981-13-6528-7_6}}


\bibitem[Selby and Woollard(2013)]%
        {selbyComputationalThinkingDeveloping}
\bibfield{author}{\bibinfo{person}{Cynthia Selby} {and} \bibinfo{person}{John
  Woollard}.} \bibinfo{year}{2013}\natexlab{}.
\newblock \bibinfo{booktitle}{\emph{Computational thinking: the developing
  definition}}.
\newblock \bibinfo{type}{Project Report}. \bibinfo{institution}{University of
  Southampton}.
\newblock
\urldef\tempurl%
\url{https://eprints.soton.ac.uk/356481/}
\showURL{%
\tempurl}


\bibitem[Selby(2015)]%
        {selbyRelationshipsComputationalThinking2015a}
\bibfield{author}{\bibinfo{person}{Cynthia~C. Selby}.}
  \bibinfo{year}{2015}\natexlab{}.
\newblock \showarticletitle{Relationships: Computational Thinking, Pedagogy of
  Programming, and {{Bloom}}'s {{Taxonomy}}}. In
  \bibinfo{booktitle}{\emph{Proceedings of the {{Workshop}} in {{Primary}} and
  {{Secondary Computing Education}}}}. \bibinfo{publisher}{ACM},
  \bibinfo{address}{London United Kingdom}, \bibinfo{pages}{80--87}.
\newblock
\showISBNx{978-1-4503-3753-3}
\href{https://doi.org/10.1145/2818314.2818315}{doi:\nolinkurl{10.1145/2818314.2818315}}


\bibitem[{\c S}en(2023)]%
        {senRelationsPreserviceTeachers2023}
\bibfield{author}{\bibinfo{person}{{\c S}enol {\c S}en}.}
  \bibinfo{year}{2023}\natexlab{}.
\newblock \showarticletitle{Relations between Preservice Teachers'
  Self-Efficacy, Computational Thinking Skills and Metacognitive
  Self-Regulation}.
\newblock \bibinfo{journal}{\emph{European Journal of Psychology of Education}}
  \bibinfo{volume}{38}, \bibinfo{number}{3} (\bibinfo{date}{Sept.}
  \bibinfo{year}{2023}), \bibinfo{pages}{1251--1269}.
\newblock
\showISSN{1878-5174}
\href{https://doi.org/10.1007/s10212-022-00651-8}{doi:\nolinkurl{10.1007/s10212-022-00651-8}}


\bibitem[Shute et~al\mbox{.}(2017)]%
        {shuteDemystifyingComputationalThinking2017}
\bibfield{author}{\bibinfo{person}{Valerie~J. Shute}, \bibinfo{person}{Chen
  Sun}, {and} \bibinfo{person}{Jodi {Asbell-Clarke}}.}
  \bibinfo{year}{2017}\natexlab{}.
\newblock \showarticletitle{Demystifying Computational Thinking}.
\newblock \bibinfo{journal}{\emph{Educational Research Review}}
  \bibinfo{volume}{22} (\bibinfo{date}{Nov.} \bibinfo{year}{2017}),
  \bibinfo{pages}{142--158}.
\newblock
\showISSN{1747-938X}
\href{https://doi.org/10.1016/j.edurev.2017.09.003}{doi:\nolinkurl{10.1016/j.edurev.2017.09.003}}


\bibitem[Soloway(1986)]%
        {solowayLearningProgramLearning1986}
\bibfield{author}{\bibinfo{person}{Elliot~M. Soloway}.}
  \bibinfo{year}{1986}\natexlab{}.
\newblock \showarticletitle{Learning to Program = Learning to Construct
  Mechanisms and Explanations}.
\newblock \bibinfo{journal}{\emph{Commun. ACM}} \bibinfo{volume}{29},
  \bibinfo{number}{9} (\bibinfo{date}{Sept.} \bibinfo{year}{1986}),
  \bibinfo{pages}{850--858}.
\newblock
\showISSN{0001-0782}
\href{https://doi.org/10.1145/6592.6594}{doi:\nolinkurl{10.1145/6592.6594}}


\bibitem[Sulistiyowati and Masduki(2024)]%
        {sulistiyowatiExplorationStudentsComputational2024}
\bibfield{author}{\bibinfo{person}{Dewi Sulistiyowati} {and}
  \bibinfo{person}{Masduki Masduki}.} \bibinfo{year}{2024}\natexlab{}.
\newblock \showarticletitle{Exploration of Students Computational Thinking
  Abilities in Solving Sequences and Series Problems Based on Learning Style}.
\newblock \bibinfo{journal}{\emph{Desimal}} \bibinfo{volume}{7},
  \bibinfo{number}{2} (\bibinfo{date}{July} \bibinfo{year}{2024}),
  \bibinfo{pages}{189--204}.
\newblock
\showISSN{2613-9081}
\href{https://doi.org/10.24042/djm.v7i2.21971}{doi:\nolinkurl{10.24042/djm.v7i2.21971}}


\bibitem[Sun et~al\mbox{.}(2024)]%
        {chendebugging2024}
\bibfield{author}{\bibinfo{person}{Chen Sun}, \bibinfo{person}{Stephanie Yang},
  {and} \bibinfo{person}{Betsy Becker}.} \bibinfo{year}{2024}\natexlab{}.
\newblock \showarticletitle{Debugging in Computational Thinking: A
  Meta-analysis on the Effects of Interventions on Debugging Skills}.
\newblock \bibinfo{journal}{\emph{Journal of Educational Computing Research}}
  \bibinfo{volume}{62}, \bibinfo{number}{4} (\bibinfo{year}{2024}),
  \bibinfo{pages}{867--901}.
\newblock
\href{https://doi.org/10.1177/07356331241227793}{doi:\nolinkurl{10.1177/07356331241227793}}


\bibitem[Weintrop et~al\mbox{.}(2016)]%
        {weintropDefiningComputationalThinking2016a}
\bibfield{author}{\bibinfo{person}{David Weintrop}, \bibinfo{person}{Elham
  Beheshti}, \bibinfo{person}{Michael Horn}, \bibinfo{person}{Kai Orton},
  \bibinfo{person}{Kemi Jona}, \bibinfo{person}{Laura Trouille}, {and}
  \bibinfo{person}{Uri Wilensky}.} \bibinfo{year}{2016}\natexlab{}.
\newblock \showarticletitle{Defining {{Computational Thinking}} for
  {{Mathematics}} and {{Science Classrooms}}}.
\newblock \bibinfo{journal}{\emph{Journal of Science Education and Technology}}
  \bibinfo{volume}{25}, \bibinfo{number}{1} (\bibinfo{year}{2016}),
  \bibinfo{pages}{127--147}.
\newblock
\showISSN{1573-1839}
\href{https://doi.org/10.1007/s10956-015-9581-5}{doi:\nolinkurl{10.1007/s10956-015-9581-5}}


\bibitem[West et~al\mbox{.}(2015)]%
        {prairielearn}
\bibfield{author}{\bibinfo{person}{Matthew West}, \bibinfo{person}{Geoffrey~L
  Herman}, {and} \bibinfo{person}{Craig Zilles}.}
  \bibinfo{year}{2015}\natexlab{}.
\newblock \showarticletitle{PrairieLearn: Mastery-based Online Problem Solving
  with Adaptive Scoring and Recommendations Driven by Machine Learning}. In
  \bibinfo{booktitle}{\emph{2015 ASEE Annual Conference \& Exposition}}.
  \bibinfo{publisher}{ASEE Conferences}, \bibinfo{address}{Seattle,
  Washington}.
\newblock
\href{https://doi.org/10.18260/p.24575}{doi:\nolinkurl{10.18260/p.24575}}


\bibitem[Wing(2006)]%
        {wingComputationalThinking2006a}
\bibfield{author}{\bibinfo{person}{Jeannette~M. Wing}.}
  \bibinfo{year}{2006}\natexlab{}.
\newblock \showarticletitle{Computational Thinking}.
\newblock \bibinfo{journal}{\emph{Commun. ACM}} \bibinfo{volume}{49},
  \bibinfo{number}{3} (\bibinfo{date}{March} \bibinfo{year}{2006}),
  \bibinfo{pages}{33--35}.
\newblock
\showISSN{0001-0782, 1557-7317}
\href{https://doi.org/10.1145/1118178.1118215}{doi:\nolinkurl{10.1145/1118178.1118215}}


\bibitem[Wing(2008)]%
        {wingComputationalThinkingThinking2008}
\bibfield{author}{\bibinfo{person}{Jeannette~M. Wing}.}
  \bibinfo{year}{2008}\natexlab{}.
\newblock \showarticletitle{Computational Thinking and Thinking about
  Computing}.
\newblock \bibinfo{journal}{\emph{Philosophical Transactions of the Royal
  Society A: Mathematical, Physical and Engineering Sciences}}
  \bibinfo{volume}{366}, \bibinfo{number}{1881} (\bibinfo{year}{2008}),
  \bibinfo{pages}{3717--3725}.
\newblock
\href{https://doi.org/10.1098/rsta.2008.0118}{doi:\nolinkurl{10.1098/rsta.2008.0118}}


\bibitem[Woo and Falloon(2022)]%
        {wooProblemSolvedHow2022}
\bibfield{author}{\bibinfo{person}{Karen Woo} {and} \bibinfo{person}{Garry
  Falloon}.} \bibinfo{year}{2022}\natexlab{}.
\newblock \showarticletitle{Problem Solved, but How? {{An}} Exploratory Study
  into Students’ Problem Solving Processes in Creative Coding Tasks}.
\newblock \bibinfo{journal}{\emph{Thinking Skills and Creativity}}
  \bibinfo{volume}{46} (\bibinfo{year}{2022}), \bibinfo{pages}{101193}.
\newblock
\showISSN{1871-1871}
\href{https://doi.org/10.1016/j.tsc.2022.101193}{doi:\nolinkurl{10.1016/j.tsc.2022.101193}}


\bibitem[Wu et~al\mbox{.}(2024)]%
        {wuIdentificationProblemSolvingTechniques2024}
\bibfield{author}{\bibinfo{person}{Ting-Ting Wu}, \bibinfo{person}{Andik
  Asmara}, \bibinfo{person}{Yueh-Min Huang}, {and} \bibinfo{person}{Intan
  Permata~Hapsari}.} \bibinfo{year}{2024}\natexlab{}.
\newblock \showarticletitle{Identification of {{Problem-Solving Techniques}} in
  {{Computational Thinking Studies}}: {{Systematic Literature Review}}}.
\newblock \bibinfo{journal}{\emph{SAGE Open}} \bibinfo{volume}{14},
  \bibinfo{number}{2} (\bibinfo{year}{2024}).
\newblock
\showISSN{2158-2440}
\href{https://doi.org/10.1177/21582440241249897}{doi:\nolinkurl{10.1177/21582440241249897}}


\bibitem[Zapata-Cáceres et~al\mbox{.}(2024)]%
        {zapatacaceresBebrasComputationalThinking2024}
\bibfield{author}{\bibinfo{person}{María Zapata-Cáceres},
  \bibinfo{person}{Pedro Marcelino}, \bibinfo{person}{Laila El-Hamamsy}, {and}
  \bibinfo{person}{Estefanía Martín-Barroso}.}
  \bibinfo{year}{2024}\natexlab{}.
\newblock \showarticletitle{A {{Bebras Computational Thinking}}
  ({{ABC-Thinking}}) Program for Primary School: {{Evaluation}} Using the
  Competent Computational Thinking Test}.
\newblock \bibinfo{journal}{\emph{Education and Information Technologies}}
  \bibinfo{volume}{29}, \bibinfo{number}{12} (\bibinfo{year}{2024}),
  \bibinfo{pages}{14969--14998}.
\newblock
\showISSN{1573-7608}
\href{https://doi.org/10.1007/s10639-023-12441-w}{doi:\nolinkurl{10.1007/s10639-023-12441-w}}


\bibitem[Zhou et~al\mbox{.}(2023)]%
        {zhouApplicationMetacognitivePlanning2023}
\bibfield{author}{\bibinfo{person}{Ying Zhou}, \bibinfo{person}{Ching~Sing
  Chai}, \bibinfo{person}{Xiuting Li}, \bibinfo{person}{Chao Ma},
  \bibinfo{person}{Baoping Li}, \bibinfo{person}{Ding Yu}, {and}
  \bibinfo{person}{Jyh-Chong Liang}.} \bibinfo{year}{2023}\natexlab{}.
\newblock \showarticletitle{Application of {{Metacognitive Planning
  Scaffolding}} for the {{Cultivation}} of {{Computational Thinking}}}.
\newblock \bibinfo{journal}{\emph{Journal of Educational Computing Research}}
  \bibinfo{volume}{61}, \bibinfo{number}{6} (\bibinfo{date}{Oct.}
  \bibinfo{year}{2023}), \bibinfo{pages}{1123--1142}.
\newblock
\showISSN{0735-6331}
\href{https://doi.org/10.1177/07356331231160294}{doi:\nolinkurl{10.1177/07356331231160294}}


\end{thebibliography}

\end{document}